\documentclass[aps,PRB,twocolumn,superscriptaddress,preprintnumbers]{revtex4-2}
\usepackage[T1]{fontenc}
\usepackage{times}
\usepackage{graphicx,color}
\usepackage{amsfonts,amsmath,amssymb,amsbsy}
\usepackage[colorlinks=true, linkcolor=blue, citecolor=blue, urlcolor=blue]{hyperref}
\usepackage{float}

\begin{document}

\title{Systematic construction of topological-nontopological hybrid
universal quantum gates \\
based on many-body Majorana fermion interactions}
\author{Motohiko Ezawa}
\affiliation{Department of Applied Physics, The University of Tokyo, 7-3-1 Hongo, Tokyo
113-8656, Japan}

\begin{abstract}
Topological quantum computation by way of braiding of Majorana fermions is
not universal quantum computation. There are several attempts to make
universal quantum computation by introducing some additional quantum gates
or quantum states. However, there is an embedding problem that $M$-qubit
gates cannot be embedded straightforwardly in $N$ qubits for $N>M$. This
problem is inherent to the Majorana system, where logical qubits are
different from physical qubits because braiding operations preserve the
fermion parity. By introducing $2N$-body interactions of Majorana fermions,
topological-nontopological hybrid universal quantum computation is shown to
be possible. Especially, we make a systematic construction of the C$^{n}$Z
gate, C$^{n}$NOT gate and the C$^{n}$SWAP gate.
\end{abstract}

\date{\today }
\maketitle

%\pacs{75.10.Hk, 75.70.Kw, 75.78.-n, 12.39.Dc}

%\email[Email:~]{ezawa@ap.t.u-tokyo.ac.jp}

%\section{Introduction}

A quantum computer is a promising next generation computer\cite%
{Feynman,DiVi,Nielsen}. In order to execute any quantum algorithms,
universal quantum computation is necessary\cite{Deutsch,Dawson,Universal}.
There are various approaches to realize universal computation including
superconductors\cite{Nakamura}, photonic systems\cite{Knill}, quantum dots%
\cite{Loss}, trapped ions\cite{Cirac} and nuclear magnetic resonance\cite%
{Vander,Kane}. The Solovay-Kitaev theorem dictates that only the Hadamard
gate, the $\pi /4$ phase-shift gate and the CNOT gate are enough for
universal quantum computation. These one and two-qubit quantum gates can be
embedded to larger qubits straightforwardly in these approaches.

Braiding of Majorana fermions is the most promising method for topological
quantum computation\cite{Brav2,Ivanov,KitaevTQC,DasTQC,TQC}. There are
various approaches to materialize Majorana fermions such as fractional
quantum Hall effects\cite{ReadGreen,ReadBraid,DasTQC,Freedman}, topological
superconductors\cite{Qi,Sato,Elli,Das,AliceaBraid,Alicea,Been} and Kitaev
spin liquids\cite{Kitaev,Matsuda}. However, it can generate only a part of
Clifford gate\cite{Brav,Ahl}. The entire Clifford gates are generated for
two qubits but not for more than three qubits\cite{Ahl}. Furthermore, only
the Clifford gates are not enough to exceed classical computers, which is
known as the Gottesman-Knill theorem\cite{GottesA,GottesB,Aaron}.

There are several attempts to make universal quantum computation based on
Majorana fermions\cite%
{Brav,Freedman,Bravi3,Sau,Das,OBrien,Bonder,Bonder2,Bark,Karzig}. In the
Majorana system, it is necessary to construct logical qubits from physical
qubits by taking a parity definite basis, because braiding preserves the
fermion parity. It makes logical qubits nonlocal. It is a nontrivial problem
to embed a nonlocal $M$-qubit quantum gate in the $N$-qubit system with $N>M$%
. Hence, even if the Hadamard gate, the $\pi /4$ phase-shift gate and the
CNOT gate are constructed, it is not enough for universal quantum
computation in the $N$-qubit system unless this embedding problem is
resolved.

In this paper, we systematically construct various quantum gates for
universal quantum computation by introducing $2N$-body interactions of
Majorana fermions preserving the fermion parity. We have required the
fermion parity preservation because it is beneficial to use the standard
braiding process as much as possible due to its topological protection. By
combining topological quantum gates generated by braiding and additional
quantum gates generated by many-body interactions of Majorana fermions,
topological-nontopological hybrid universal quantum computation is possible.
It would be more robust than conventional universal quantum computation
because the quantum gates generated by braiding are topologically protected.
We systematically construct arbitrary C$^{n}$-phase shift gates, the
Hadamard gate, C$^{n}$NOT gates and C$^{n}$SWAP gates in the $N$-qubit
system by this generalization.

Supplementary Materials are prepared for detailed analysis in the case of
small qubits to make clear a general analysis for the $N$-qubit system.

\textbf{Physical qubits and logical qubits:} Majorana fermions are described
by operators $\gamma _{\alpha }$\ satisfying the anticommutation relations%
\begin{equation}
\left\{ \gamma _{\alpha },\gamma _{\beta }\right\} =2\delta _{\alpha \beta }.
\end{equation}%
The braid operator is defined by\cite{Ivanov}%
\begin{equation}
\mathcal{B}_{\alpha \beta }=\exp \left[ \frac{\pi }{4}\gamma _{\beta }\gamma
_{\alpha }\right] =\frac{1}{\sqrt{2}}\left( 1+\gamma _{\beta }\gamma
_{\alpha }\right) .
\end{equation}%
It satisfies $\mathcal{B}_{\alpha \beta }^{4}=1$ and there is a
corresponding anti-braiding operator $\mathcal{B}_{\alpha \beta }^{-1}=%
\mathcal{B}_{\alpha \beta }^{3}$.

The qubit basis is defined by\cite{Ivanov}%
\begin{eqnarray}
&&\left\vert n_{N+1}n_{N}\cdots n_{2}n_{1}\right\rangle _{\text{physical}} 
\notag \\
&\equiv &\left( c_{1}^{\dagger }\right) ^{n_{1}}\left( c_{2}^{\dagger
}\right) ^{n_{2}}\cdots \left( c_{N}^{\dagger }\right) ^{n_{N}}\left(
c_{N+1}^{\dagger }\right) ^{n_{N+1}}\left\vert 0\right\rangle ,
\end{eqnarray}%
with $n_{\alpha }=0$ or $1$, where ordinary fermion operators are
constructed from two Majorana fermions as 
\begin{equation}
c_{\alpha }=\frac{1}{2}\left( \gamma _{2\alpha -1}+i\gamma _{2\alpha
}\right) .
\end{equation}%
$2N+4$ Majorana fermions constitute $N+1$ physical qubits.

The braiding operation preserves the fermion parity $P_{\alpha \beta }\equiv
i\gamma _{\beta }\gamma _{\alpha }$, where it commutes with the braid
operator $\mathcal{B}_{\alpha \beta }$, 
\begin{equation}
\left[ \mathcal{B}_{\alpha \beta },P_{\alpha \beta }\right] =0.
\end{equation}%
It means that if we start with the even parity state $\left\vert 00\cdots
0\right\rangle _{\text{physical}}$, the states after any braiding process
should have even fermion parity. Therefore, in order to construct $N$
logical qubits $\left\vert n_{N}\cdots n_{2}n_{1}\right\rangle _{\text{%
logical}}$, $N+1$ physical qubits $\left\vert n_{N+1}^{\prime }\cdots
n_{2}^{\prime }n_{1}^{\prime }\right\rangle _{\text{physical}}$ are necessary%
\cite{NayakWilczek,GeorgievJS,GeorgievJP}. There are $N!$ correspondences
between the logical and physical qubits in general. However, we adopt the
following unique correspondence. When the logical qubit $\left\vert
n_{N}\cdots n_{2}n_{1}\right\rangle _{\text{logical}}$ is given, we
associate to it a physical qubit $\left\vert n_{N}\cdots
n_{2}n_{1}n_{0}\right\rangle _{\text{physical}}$ by adding one qubit $n_{0}$
uniquely so that $\sum_{\alpha =0}^{N}n_{\alpha }=0$ mod $2$. Alternatively,
when a physical qubit $\left\vert n_{N}\cdots n_{2}n_{1}n_{0}\right\rangle _{%
\text{logical}}$ is given, we associate to it a logical qubit $\left\vert
n_{N}\cdots n_{2}n_{1}\right\rangle _{\text{physical}}$ just by eliminating
the qubit $n_{0}$. An example reads as follows,%
\begin{equation}
\left( \overbrace{%
\begin{array}{c}
\left\vert 0,\cdots ,0,0,0\right\rangle \\ 
\left\vert 0,\cdots ,0,0,1\right\rangle \\ 
\left\vert 0,\cdots ,0,1,0\right\rangle \\ 
\left\vert 0,\cdots ,0,1,1\right\rangle \\ 
\left\vert 0,\cdots ,1,0,0\right\rangle \\ 
\left\vert 0,\cdots ,1,0,1\right\rangle \\ 
\cdots%
\end{array}%
}^{N}\right) _{\text{logical}}=\left( \overbrace{%
\begin{array}{c}
\left\vert 0,\cdots ,0,0,0,0\right\rangle \\ 
\left\vert 0,\cdots ,0,0,1,1\right\rangle \\ 
\left\vert 0,\cdots ,0,1,0,1\right\rangle \\ 
\left\vert 0,\cdots ,0,1,1,0\right\rangle \\ 
\left\vert 0,\cdots ,1,0,0,1\right\rangle \\ 
\left\vert 0,\cdots ,1,0,1,0\right\rangle \\ 
\cdots%
\end{array}%
}^{N+1}\right) _{\text{physical}}.  \label{LogiPhys}
\end{equation}%
This correspondence is different from those in the previous works\cite%
{Brav2,GeorgievJS,GeorgievJP,GeorgievB,GeorgievNPB,Kraus}. Accordingly, the
detailed braiding process for quantum gates are slightly different from the
previous ones\cite{GeorgievJS,GeorgievJP,GeorgievB,GeorgievNPB,Kraus}.

\textbf{$2N$-body interactions:} 
%\subsection{$2$-body interaction of Majorana fermions}
A generic operator involving two Majorana fermions $\gamma _{\alpha }$ and $%
\gamma _{\beta }$ is expressed as $U_{\alpha \beta }=a_{1}+a_{2}\gamma
_{\alpha }+a_{3}\gamma _{\beta }+a_{4}\gamma _{\beta }\gamma _{\alpha }$,
since higher-order terms $\gamma _{\alpha }^{p}\gamma _{\beta }^{q}$ are
absent for $p\geq 2$ and $q\geq 2$ because of the relations $\gamma _{\alpha
}^{2}=\gamma _{\beta }^{2}=1$. Then, by imposing the parity conservation
condition $\left[ U_{\alpha \beta },P_{\alpha \beta }\right] =0$ with the
fermion-parity operator $P_{\alpha \beta }=\gamma _{\beta }\gamma \alpha $,
it is restricted to $U_{\alpha \beta }=a_{1}+a_{4}\gamma _{\beta }\gamma
_{\alpha }$. Furthermore, the unitary condition $U_{\alpha \beta }U_{\alpha
\beta }=U_{\alpha \beta }^{\dagger }U_{\alpha \beta }=1$ leads to the
representation of $U_{\alpha \beta }$ in the form of%
\begin{align}
U_{\alpha \beta }\left( \theta \right) & =\exp (\theta \gamma _{\beta
}\gamma _{\alpha })=\cos \theta +\gamma _{\beta }\gamma _{\alpha }\sin
\theta ,  \label{Uij} \\
(U_{\alpha \beta }\left( \theta \right) )^{-1}& =\exp (-\theta \gamma
_{\beta }\gamma _{\alpha })=\cos \theta -\gamma _{\beta }\gamma _{\alpha
}\sin \theta .
\end{align}%
The choice $\theta =\pi /4$ corresponds to the braiding operation. In
general, $\theta $ can take an arbitrary value.

This operator transforms the Majorana operators as%
\begin{align}
U_{\alpha \beta }\left( \theta \right) \gamma _{\alpha }(U_{\alpha \beta
}\left( \theta \right) )^{-1}& =\gamma _{\alpha }\cos 2\theta +\gamma
_{\beta }\sin 2\theta ,  \notag \\
U_{\alpha \beta }\left( \theta \right) \gamma _{\beta }(U_{\alpha \beta
}\left( \theta \right) )^{-1}& =-\gamma _{\alpha }\sin 2\theta +\gamma
_{\beta }\cos 2\theta .  \label{PartialB}
\end{align}%
We show that non-Clifford gates are constructed based on them.

The two-body operation is realized by the unitary dynamics,%
\begin{equation}
\mathcal{B}_{\alpha \beta }\left( \theta \right) =\exp \left[ \theta \gamma
_{\beta }\gamma _{\alpha }\right] =\exp \left[ iHt/\hbar \right] ,
\end{equation}%
with%
\begin{equation}
H=\frac{\hbar \theta }{t}\gamma _{\beta }\gamma _{\alpha }.
\end{equation}

The four-body operation%
\begin{equation}
\mathcal{B}_{1234}^{(4)}\equiv \exp \left[ i\frac{\pi }{4}\gamma _{4}\gamma
_{3}\gamma _{2}\gamma _{1}\right]
\end{equation}%
is introduced\cite{Brav2} as an essential ingredient of universal quantum
computation. We generalize it to the four-body operation $\mathcal{B}%
_{\alpha \beta \gamma \delta }^{\left( 4\right) }\left( \theta \right) $
defined by%
\begin{eqnarray}
\mathcal{B}_{\alpha \beta \gamma \delta }^{(4)}\left( \theta \right) &\equiv
&\exp \left[ i\theta \gamma _{\delta }\gamma _{\gamma }\gamma _{\beta
}\gamma _{\alpha }\right]  \notag \\
&=&\cos \theta +i\gamma _{\delta }\gamma _{\gamma }\gamma _{\beta }\gamma
_{\alpha }\sin \theta ,  \label{B4}
\end{eqnarray}%
which keeps the parity%
\begin{equation}
\left[ \mathcal{B}_{\alpha \beta \gamma \delta }^{\left( 4\right) }\left(
\theta \right) ,P_{\alpha \beta }\right] =0.
\end{equation}%
We use an abbreviation,%
\begin{equation}
\mathcal{B}_{\alpha }^{\left( 4\right) }\left( \theta \right) \equiv 
\mathcal{B}_{\alpha ,\alpha +1.\alpha +2,\alpha +3}^{(4)}\left( \theta
\right) .
\end{equation}%
The relation 
\begin{equation}
\left( \mathcal{B}_{1}^{\left( 4\right) }\left( \theta \right) \right)
^{\dagger }\mathcal{B}_{1}^{\left( 4\right) }\left( \theta \right) =I
\end{equation}%
holds because of the coefficient $i$ in Eq.(\ref{B4}).

It is realized by the unitary dynamics,%
\begin{equation}
\mathcal{B}_{\alpha \beta \gamma \delta }^{\left( 4\right) }\left( \theta
\right) =\exp \left[ iHt/\hbar \right] ,
\end{equation}%
with%
\begin{equation}
H=\frac{i\hbar \theta }{t}\gamma _{\delta }\gamma _{\gamma }\gamma _{\beta
}\gamma _{\alpha }.
\end{equation}

Similary, we may define the $2N$-body operation by%
\begin{eqnarray}
\mathcal{B}_{1}^{\left( 2N\right) }\left( \theta \right) &\equiv &\mathcal{B}%
_{1\sim 2N}^{(2N)}\left( \theta \right) \equiv \exp \left[ i^{(N-1)\theta
}\gamma _{2N}\gamma _{2N-1}\cdots \gamma _{2}\gamma _{1}\right]  \notag \\
&=&\cos \theta +i^{N-1}\gamma _{2N}\gamma _{2N-1}\cdots \gamma _{2}\gamma
_{1}\sin \theta .
\end{eqnarray}%
It satisfies the unitary condition,%
\begin{equation}
\left( \mathcal{B}_{1}^{\left( 2N\right) }\left( \theta \right) \right)
^{\dagger }\mathcal{B}_{1}^{\left( 2N\right) }\left( \theta \right) =I.
\end{equation}%
We also define%
\begin{equation}
\mathcal{B}_{\alpha }^{\left( 2N\right) }\left( \theta \right) \equiv 
\mathcal{B}_{\alpha ,\alpha +1.\cdots \alpha +2N-2,\alpha +2N-1}\left(
\theta \right) .
\end{equation}%
It is realized by the $2N$-body interaction of Majorana fermions%
\begin{equation}
H=\frac{i^{N-1}\hbar \theta }{t}\gamma _{2N}\gamma _{2N-1}\cdots \gamma
_{2}\gamma _{1}.
\end{equation}

\textbf{$N$ physical qubits: } We consider the $2N$ Majorana fermion system.
The explicit actions on $2N$ physical qubits are given by%
\begin{eqnarray}
\mathcal{B}_{1}\left( \theta \right) &=&I_{2N-2}\otimes R_{z}\left( 2\theta
\right) ,  \notag \\
\mathcal{B}_{3}\left( \theta \right) &=&I_{2N-4}\otimes R_{z}\left( 2\theta
\right) \otimes I_{2},  \notag \\
&&\cdots  \notag \\
\mathcal{B}_{2n-1}\left( \theta \right) &=&I_{2N-2n}\otimes R_{z}\left(
2\theta \right) \otimes I_{2n-2},  \notag \\
&&\cdots  \notag \\
\mathcal{B}_{2N-1}\left( \theta \right) &=&R_{z}\left( 2\theta \right)
\otimes I_{2N-2}
\end{eqnarray}%
for odd numbers and%
\begin{align}
\mathcal{B}_{2}\left( \theta \right) & =I_{2N-4}\otimes U_{xx}\left( \theta
\right) ,  \notag \\
& \cdots  \notag \\
\mathcal{B}_{2n}\left( \theta \right) & =I_{2N-2-2n}\otimes U_{xx}\left(
\theta \right) \otimes I_{2n-2},  \notag \\
& \cdots  \notag \\
\mathcal{B}_{2N-2}\left( \theta \right) & =U_{xx}\left( \theta \right)
\otimes I_{2N-4}  \label{BPauliX}
\end{align}%
for even numbers, where we have defined the rotation along the $z$ axis by%
\begin{equation}
R_{z}\left( \theta \right) \equiv \exp \left[ -i\frac{\theta }{2}\sigma _{z}%
\right] =\text{diag.}\left( e^{-i\theta /2},e^{i\theta /2}\right) ,
\end{equation}%
and%
\begin{equation}
U_{xx}\left( \theta \right) \equiv \exp \left[ -i\theta \sigma _{x}\otimes
\sigma _{x}\right] .
\end{equation}

\textbf{$N-1$ logical qubits:} $N-1$ logical qubits are constructed from $N$
physical qubits based on the correspondence (\ref{LogiPhys}).

Local $z$ rotation is possible for any qubits as in%
\begin{eqnarray}
\mathcal{B}_{1}\left( \theta \right)  &=&\exp \left[ -i\theta
\bigotimes_{j=1}^{N-1}\sigma _{z}\right] ,  \notag \\
\mathcal{B}_{3}\left( \theta \right)  &=&I_{2N-4}\otimes R_{z}\left( 2\theta
\right)   \notag \\
&=&\exp \left[ -i\theta I_{2}\otimes I_{2}\otimes \sigma _{z}\right] , 
\notag \\
&&\cdots   \notag \\
\mathcal{B}_{2n-1}\left( \theta \right)  &=&I_{2N-2n}\otimes R_{z}\left(
2\theta \right) \otimes I_{2n-4}  \notag \\
&=&\exp \left[ -i\theta I_{2N-2n}\otimes \sigma _{z}\otimes I_{2n-4}\right] ,
\notag \\
&&\cdots   \notag \\
\mathcal{B}_{2N-3}\left( \theta \right)  &=&I_{2}\otimes R_{z}\left( 2\theta
\right) \otimes I_{2N-2n-6}  \notag \\
&=&\exp \left[ -i\theta I_{2}\otimes \sigma _{z}\otimes I_{2N-2n-6}\right] .
\end{eqnarray}%
Local $x$ rotation is possible for any qubits as in%
\begin{eqnarray}
\mathcal{B}_{23}\left( \theta \right)  &=&I_{2N-4}\otimes R_{x}\left(
2\theta \right) ,  \notag \\
\mathcal{B}_{2\sim 5}^{\left( 4\right) }\left( \theta \right) 
&=&I_{2N-6}\otimes R_{x}\left( 2\theta \right) \otimes I_{2},  \notag \\
\mathcal{B}_{2\sim 7}^{\left( 6\right) }\left( \theta \right) 
&=&I_{2N-8}\otimes R_{x}\left( 2\theta \right) \otimes I_{4},  \notag \\
&&\cdots   \notag \\
\mathcal{B}_{2\sim 2N-3}^{\left( 2N-4\right) }\left( \theta \right) 
&=&I_{2}\otimes R_{x}\left( 2\theta \right) \otimes I_{2N-6},  \notag \\
\mathcal{B}_{2\sim 2N-1}^{\left( 2N-2\right) }\left( \theta \right) 
&=&R_{x}\left( 2\theta \right) \otimes I_{2N-4},
\end{eqnarray}%
where we have defined the rotation along the $x$ axis by%
\begin{equation}
R_{x}\left( \theta \right) \equiv \exp \left[ -i\frac{\theta }{2}\sigma _{x}%
\right] =\left( 
\begin{array}{cc}
\cos \frac{\theta }{2} & -i\sin \frac{\theta }{2} \\ 
-i\sin \frac{\theta }{2} & \cos \frac{\theta }{2}%
\end{array}%
\right) ,
\end{equation}%
and $\mathcal{B}_{\alpha \sim \beta }\left( \theta \right) \equiv \mathcal{B}%
_{\alpha ,\alpha +1,\cdots \beta -1,\beta }\left( \theta \right) $.
Accordingly, the Hadamard gate is embedded in arbitrary $N$ qubits by using
the decomposition formula,%
\begin{equation}
U_{\text{H}}=R_{z}\left( \frac{\pi }{4}\right) R_{x}\left( \frac{\pi }{4}%
\right) R_{z}\left( \frac{\pi }{4}\right) .
\end{equation}

Next, we show that it is possible to construct any $2^{N}$ diagonal
operators based on many-body Majorana interactions, and hence, an arbitrary C%
$^{n}$-phase-shift gate is constructed. By applying the Hadamard gate, it is
possible to construct the C$^{n}$NOT gate. There are $_{N}C_{M}$ patterns of 
$2M$-body unitary evolutions in $2N$ physical qubits. By taking a sum, we
have $\sum_{M=1N}^{N}C_{M}=2^{N}$ independent physical qubits. They produce $%
2^{N-1}$ independent logical qubits because there are complementary
operators $\mathcal{B}_{\alpha }\left( \theta \right) $ and $\mathcal{B}_{%
\overline{\alpha }}\left( \theta \right) $ which produce the same logical
qubits. The complementary operators are denoted as%
\begin{equation}
\mathcal{B}_{\overline{\alpha }}\left( \theta \right) \simeq \mathcal{B}%
_{\alpha }\left( \theta \right) ,
\end{equation}%
where $\overline{\alpha }$\ indicates a set of indices which is the
complementary set of $\alpha $\ in the indices $1\sim 2N$. For example, we
consider the case for four qubits $N=4$, where eight Majorana fermions
exists. The following different braiding operators $\mathcal{B}_{56}$ and $%
\mathcal{B}_{123478}$\ give an identical logical quantum gate%
\begin{equation}
\mathcal{B}_{123478}\simeq \mathcal{B}_{56},
\end{equation}%
where $\alpha =56$ and $\overline{\alpha }=123478$ with $N=4$. See the full
list of the complementary braiding operators for four physical qubits in Eq.(%
\ref{CompB}) in Supplementary Material.

There are $2^{N-1}$ independent components in $N-1$ logical qubits. On the
other hand, there are $2^{N-1}$ independent many-body Majorana operators.
Hence, it is possible to construct arbitrary diagonal operators by solving
the linear equation. They include the C$^{n}$-phase shift gates. Using the
relation%
\begin{equation}
U_{\text{C}^{n}\phi }=e^{\frac{i\phi }{2^{n+1}}}\prod_{q=1}^{{2}%
^{n+1}-1}\exp \left[ \frac{\left( -1\right) ^{\text{Mod}_{2}%
\sum_{p=1}^{n+1}p_{n}}}{2^{n+1}}\bigotimes_{p=1}^{n+1}\left( \sigma
_{z}\right) ^{q_{p}}\right] 
\end{equation}%
with $q_{p}=0,1$, the C$^{n}$-phase shift gate is constructed as%
\begin{equation}
U_{\text{C}^{n}\phi }=e^{\frac{i\phi }{2^{n+1}}}\prod_{q=1}^{{2}^{n}}%
\mathcal{B}_{\text{odd},q}\left( \frac{\phi }{2^{n+1}}\right) \prod_{r=1}^{{2%
}^{n}-1}\mathcal{B}_{\text{even},r}\left( -\frac{\phi }{2^{n+1}}\right) ,
\end{equation}%
where $\mathcal{B}_{\text{odd}}$ contains odd number of $\sigma _{z}$
operators for logical qubits, while $\mathcal{B}_{\text{even}}$ contains
even number of $\sigma _{z}$ operators for logical qubits. By setting $\phi
=\pi $, we obtain the C$^{n}$Z gate.

\begin{figure}[t]
\centerline{\includegraphics[width=0.48\textwidth]{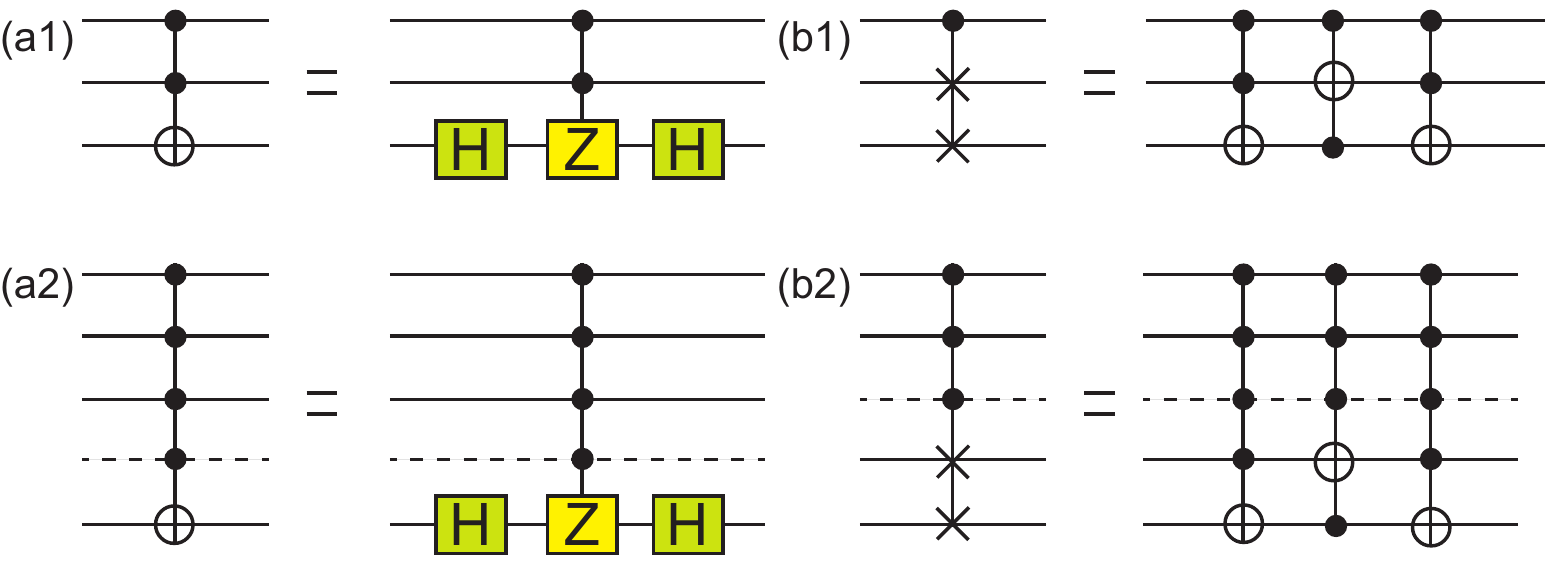}}
\caption{(a1) Construction of the CCNOT gate from the CCZ gate and the
Hadamard gates. (a2) Construction of the C$^{n}$NOT gate from the C$^{n}$Z
gate and the Hadamard gates. (b1) Construction of the Fredkin gate from
three Toffoli gates. (b2) Construction of the C$^{n}$SWAP gate from the C$%
^{n}$NOT gates.}
\label{Fredkin}
\end{figure}

For example, the CZ gate in three qubits are embedded as%
\begin{align}
U_{\text{CZ}}^{3\rightarrow 2}& =U_{\text{CZ}}\otimes I_{2}=e^{i\pi /4}%
\mathcal{B}_{56}\left( \frac{\pi }{4}\right) \mathcal{B}_{78}\left( \frac{%
\pi }{4}\right) \mathcal{B}_{1234}^{\left( 4\right) }\left( -\frac{\pi }{4}%
\right) ,  \notag \\
U_{\text{CZ}}^{3\rightarrow 1}& =e^{i\pi /4}\mathcal{B}_{34}\left( \frac{\pi 
}{4}\right) \mathcal{B}_{78}\left( \frac{\pi }{4}\right) \mathcal{B}%
_{1256}^{\left( 4\right) }\left( -\frac{\pi }{4}\right) ,  \notag \\
U_{\text{CZ}}^{2\rightarrow 1}& =I_{2}\otimes U_{\text{CZ}}=e^{i\pi /4}%
\mathcal{B}_{34}\left( \frac{\pi }{4}\right) \mathcal{B}_{56}\left( \frac{%
\pi }{4}\right) \mathcal{B}_{1278}^{\left( 4\right) }\left( -\frac{\pi }{4}%
\right) ,  \label{EmbedCZ}
\end{align}%
where $U_{\text{CZ}}^{p\rightarrow q}$ indicates that the controlloed qubit
is $p$ and the target qubit is $q$. The CC$\phi $ phase-shift gate acting on
three logical qubits in given by%
\begin{eqnarray}
U_{\text{CC}\phi } &=&e^{i\phi /8}\mathcal{B}_{12}\left( \frac{\phi }{8}%
\right) \mathcal{B}_{34}\left( \frac{\phi }{8}\right) \mathcal{B}_{56}\left( 
\frac{\phi }{8}\right) \mathcal{B}_{78}\left( \frac{\phi }{8}\right)   \notag
\\
&&\mathcal{B}_{1234}^{\left( 4\right) }\left( -\frac{\phi }{8}\right) 
\mathcal{B}_{1278}^{\left( 4\right) }\left( -\frac{\phi }{8}\right) \mathcal{%
B}_{1256}^{\left( 4\right) }\left( -\frac{\phi }{8}\right) .
\end{eqnarray}%
Especially, the CCZ gate is constructed as follows,%
\begin{eqnarray}
U_{\text{CCZ}} &=&e^{i\pi /8}\mathcal{B}_{12}\left( \frac{\pi }{8}\right) 
\mathcal{B}_{34}\left( \frac{\pi }{8}\right) \mathcal{B}_{56}\left( \frac{%
\pi }{8}\right) \mathcal{B}_{78}\left( \frac{\pi }{8}\right)   \notag \\
&&\mathcal{B}_{1234}^{\left( 4\right) }\left( -\frac{\pi }{8}\right) 
\mathcal{B}_{1256}^{\left( 4\right) }\left( -\frac{\pi }{8}\right) \mathcal{B%
}_{1278}^{\left( 4\right) }\left( -\frac{\pi }{8}\right) .
\end{eqnarray}%
The Toffoli gate is constructed by applying the Hadamard gate to the CCZ
gate as in%
\begin{equation}
U_{\text{Toffloi}}=\left( I_{4}\otimes U_{\text{H}}\right) U_{\text{CCZ}%
}\left( I_{4}\otimes U_{\text{H}}\right) .
\end{equation}%
See Fig.\ref{Fredkin}(a1). The Fredkin gate is constructed by sequential
applications of three Toffoli gates as in%
\begin{equation}
U_{\text{Fredkin}}=U_{\text{Toffoli}}^{\left( 3,2\right) \rightarrow 1}U_{%
\text{Toffoli}}^{\left( 3,1\right) \rightarrow 2}U_{\text{Toffoli}}^{\left(
3,2\right) \rightarrow 1},
\end{equation}%
where $U_{\text{CZ}}^{\left( p,q\right) \rightarrow r}$ indicates that the
controlloed qubits are $p$ and $q$ while the target qubit is $r$. See Fig.%
\ref{Fredkin}(b1).

The C$^{n}$NOT gate is constructed from C$^{n}$Z gate as%
\begin{equation}
U_{\text{C}^{n}\text{NOT}}=\left( I_{2n-2}\otimes U_{\text{H}}\right) U_{%
\text{C}^{n}\text{Z}}\left( I_{2n-2}\otimes U_{\text{H}}\right) ,
\end{equation}%
where the Hadamard gate is applied to $n$th qubit. See Fig.\ref{Fredkin}(a2).

The C$^{n}$SWAP gate is constructed from the C$^{n}$Z gate as%
\begin{equation}
U_{\text{C}^{n}\text{SWAP}}=U_{\text{C}^{n}\text{NOT}}^{\overline{1}%
\rightarrow 1}U_{\text{C}^{n}\text{NOT}}^{\overline{2}\rightarrow 2}U_{\text{%
C}^{n}\text{NOT}}^{\overline{1}\rightarrow 1},
\end{equation}%
where $U_{\text{C}^{n}\text{NOT}}^{\overline{p}\rightarrow p}$\ indicates
that the target qubit is $p$\ and the others are controlled qubits, where $%
\overline{p}$\ indicates the complementary qubits of the qubit $p$. See Fig.%
\ref{Fredkin}(b2).

As a result, an arbitrary phase-shift gate, the C$^{n}$NOT gate and the
Hadamard gate are constructed in any logical qubits, and hence, the
universal quantum computation is possible based on many-body interactions.

\textbf{Discussions:} We have analyzed the embedding problem inherent to the
Majorana system, and shown that universal quantum computation is possible by
introducing many-body interactions of Majorana fermions. Especially, the C$%
^{n}$-phase shift gate, the C$^{n}$NOT and the C$^{n}$SWAP gate are
systematically constructed, which are the basic ingredients of universal
quantum computation based on the Solovay-Kitaev theorem. Although it was
previously pointed out that four-body interactions of Majorana fermions are
enough for universal quantum computation\cite{Brav2}, we have shown that the
four-body interactions are not enough but $2N$-body interactions are
necessary.

The proposed quantum gates based on many-body interactions of Majorana
fermions are not topologically protected because they are not generated by
the standard braiding operations. By combining topological quantum
computation based on braiding of Majorana fermions and nontopological
quantum computation based on many-body interaction of Majorana fermions,
topological-nontopological hybrid universal quantum computation is possible.
It would be more robust than conventional universal quantum computation
because many of quantum gates generated by braiding are topologically
protected.

Recently, quantum simulation on Majorana fermions is studied in
superconducting qubits\cite{Huang,Harle,Rancic}. It will be possible to
realize many-body interactions of Majorana fermions in near future.

%%%%%%%%%%%%%%%%%%%%%%%%%%%%%%%%%%%%%%%%%%%%%%%%%%%%%%%%%%%%
This work is supported by CREST, JST (Grants No. JPMJCR20T2) and
Grants-in-Aid for Scientific Research from MEXT KAKENHI (Grant No.
23H00171). %%%%%%%%%%%%%%%%%%%%%%%%%%%%%%%%%%%%%%%%%%%%%%%%%%%%%%%%%%%%

%%%%%%%%%%%%%%%
\clearpage\newpage
\onecolumngrid
\def\theequation{S\arabic{equation}}
\def\thefigure{S\arabic{figure}}
\def\thesubsection{S\arabic{subsection}}
\setcounter{figure}{0}
\setcounter{equation}{0}
\setcounter{section}{0}

\begin{center}
\textbf{\Large Supplemental Material} \bigskip \bigskip

\textbf{\large Systematic construction of topological-nontopological hybrid
universal quantum gates \\[0pt]
based on many-body Majorana fermion interactions }\bigskip

{Motohiko Ezawa}

{Department of Applied Physics, The University of Tokyo, 7-3-1 Hongo, Tokyo
113-8656, Japan}
\end{center}

\bigskip

\section{Results on conventional braiding}

\subsection{Embedding}

We consider a one-dimensional chain of Majorana fermions and only consider
the braiding between adjacent Majorana fermions. We denote $\mathcal{B}%
_{\alpha }\equiv \mathcal{B}_{\alpha ,\alpha +1}$. The braid operators $%
\mathcal{B}_{\alpha }$ satisfies the Artin braid group relation\cite{ArtinS}%
\begin{align}
\mathcal{B}_{\alpha }\mathcal{B}_{\beta }& =\mathcal{B}_{\beta }\mathcal{B}%
_{\alpha }\qquad \text{for\quad }\left\vert \alpha -\beta \right\vert \geq 2,
\notag \\
\mathcal{B}_{\alpha }\mathcal{B}_{\alpha +1}\mathcal{B}_{\alpha }& =\mathcal{%
B}_{\alpha +1}\mathcal{B}_{\alpha }\mathcal{B}_{\alpha +1}.
\end{align}

The embedding of an $M$-qubit quantum gate to an $N$-qubit system with $M<N $
is a nontrivial problem in braiding of Majorana fermions. There are two
partial solutions. One is setting additional qubits to be $0$ as ancilla
qubits, where every quantum gates can be embedded. The other is not to use
the braiding $\mathcal{B}_{1}$. We discuss both of these in what follows.

\subsubsection{Ancilla embedding}

$N-1$ logical qubits are embedded in $N$ logical qubits if the additional
qubit is $0$,%
\begin{equation}
\left\vert 0n_{N-1}\cdots n_{2}n_{1}\right\rangle _{\text{logical}}.
\end{equation}%
This is because the correspondence between the physical and logical qubits
are identical if the $N$-th qubit is $0$. It is assured by the fact that we
can use the same even parity basis in $(N-1)$ qubits because the $N$th qubit
is $0$. On the other hand, the action is different if the additional qubit
is $1$,%
\begin{equation}
\left\vert 1n_{N-1}\cdots n_{2}n_{1}\right\rangle _{\text{logical}}.
\end{equation}%
This is because it is necessary to use the odd parity basis in physical $N-1$
qubits so that total parity is even in the presence of the $N$th qubit. It
is still useful because there are many quantum algorithms where ancilla
qubits are $0$.

\subsubsection{Braid construction}

We study what $M$-qubit quantum gates can be embedded to an $N$-qubit
quantum gate with $M<N$. First, we examine the case for one logical qubit as
a simplest example. The braiding $\mathcal{B}_{1}$ acts differently on the
even and odd bases,%
\begin{equation}
\mathcal{B}_{1}^{\text{even}}\neq \mathcal{B}_{1}^{\text{odd}},
\end{equation}%
where%
\begin{equation}
\mathcal{B}_{1}^{\text{even}}=e^{-i\pi /4}\left( 
\begin{array}{cc}
1 & 0 \\ 
0 & i%
\end{array}%
\right) ,\qquad \mathcal{B}_{1}^{\text{odd}}=e^{-i\pi /4}\left( 
\begin{array}{cc}
i & 0 \\ 
0 & 1%
\end{array}%
\right) .
\end{equation}%
On the other hand, $\mathcal{B}_{2}$ and $\mathcal{B}_{3}$ act on even and
odd bases in the same way,%
\begin{equation}
\mathcal{B}_{2}^{\text{even}}=\mathcal{B}_{2}^{\text{odd}}=\frac{1}{\sqrt{2}}%
\left( 
\begin{array}{cc}
1 & -i \\ 
-i & 1%
\end{array}%
\right) ,\qquad \mathcal{B}_{3}^{\text{even}}=\mathcal{B}_{3}^{\text{odd}%
}=e^{-i\pi /4}\left( 
\begin{array}{cc}
1 & 0 \\ 
0 & i%
\end{array}%
\right) .
\end{equation}%
Similarly, $\mathcal{B}_{k}$ for $k\geq 4$ has the same action on the even
and odd bases. We find that embedding is possible if we do not use the
braiding $\mathcal{B}_{1}$. Hence, all of the Pauli gates, the Hadamard
transformation, the $i$SWAP gate can be embedded to the $N$-qubit quantum
gates.

On the other hand, the quantum gates which use braiding $\mathcal{B}_{1}$
cannot be embedded to larger qubit as it is. For example, the CZ gate is
given by the braiding $e^{-i\pi /4}\mathcal{B}_{5}^{-1}\left( \mathcal{B}%
_{3}\right) ^{-1}\mathcal{B}_{1}$, whose matrix representation is%
\begin{equation}
\text{diag.}\left( 1,1,1,-1,i,-i,-i,-i\right) ,
\end{equation}%
once it is embedded to three logical qubits. They are different, 
\begin{equation}
e^{-i\pi /4}\mathcal{B}_{5}^{-1}\left( \mathcal{B}_{3}\right) ^{-1}\mathcal{B%
}_{1}\neq I_{2}\otimes U_{\text{CZ}}=\text{diag.}\left(
1,1,1,-1,1,1,1,-1\right) ,
\end{equation}%
In general, $M$-quantum gates cannot be embedded in $N$ qubit. We solve the
problem by introducing many-body interaction of Majorana fermions in Eq.(\ref%
{EmbedCZ}).

\subsection{Single logical qubit}

We discuss how to construct single logical qubit\cite{IvanovS}. Two ordinary
fermions $c_{1}$ and $c_{2}$ are introduced from four Majorana fermions as%
\begin{equation}
c_{1}=\frac{1}{2}\left( \gamma _{1}+i\gamma _{2}\right) ,\qquad c_{2}=\frac{1%
}{2}\left( \gamma _{3}+i\gamma _{4}\right) .
\end{equation}%
The basis of physical qubits is given by%
\begin{equation}
\Psi _{\text{physical}}=\left( \left\vert 0\right\rangle ,c_{1}^{\dagger
}\left\vert 0\right\rangle ,c_{2}^{\dagger }\left\vert 0\right\rangle
,c_{1}^{\dagger }c_{2}^{\dagger }\left\vert 0\right\rangle \right)
^{t}\equiv \left( \left\vert 0,0\right\rangle _{\text{physical}},\left\vert
0,1\right\rangle _{\text{physical}},\left\vert 1,0\right\rangle _{\text{%
physical}},\left\vert 1,1\right\rangle _{\text{physical}}\right) ^{t}.
\label{PsiTwo}
\end{equation}%
By taking the even parity basis as%
\begin{equation}
\left( 
\begin{array}{c}
\left\vert 0\right\rangle \\ 
\left\vert 1\right\rangle%
\end{array}%
\right) _{\text{logical}}=\left( 
\begin{array}{c}
\left\vert 0,0\right\rangle \\ 
\left\vert 1,1\right\rangle%
\end{array}%
\right) _{\text{physical}},
\end{equation}%
single logical qubit is constructed from two physical qubits.

\subsection{Quantum gates for one logical qubit}

\begin{figure}[t]
\centerline{\includegraphics[width=0.68\textwidth]{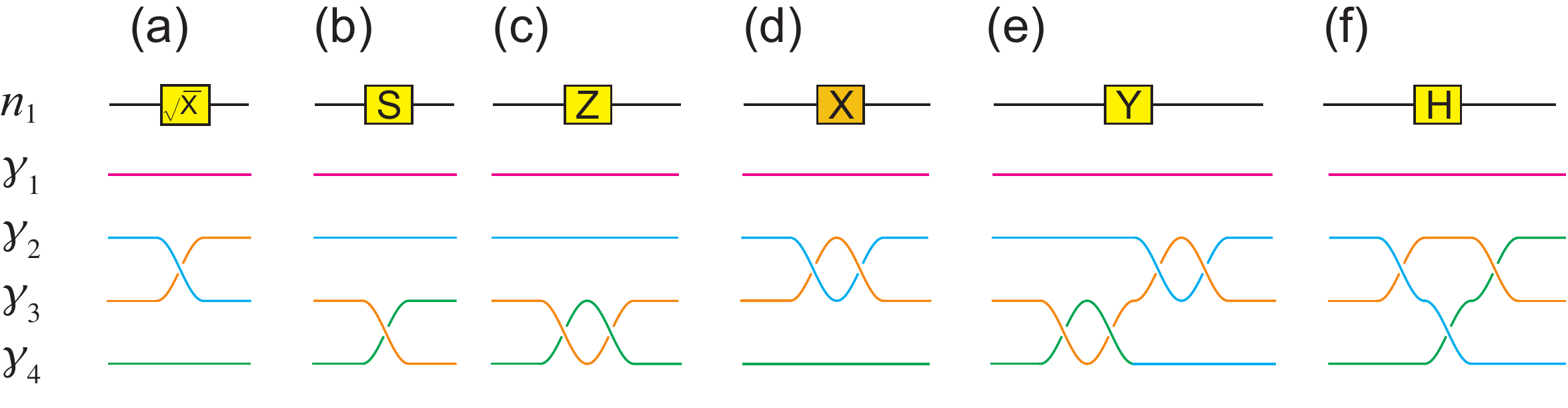}}
\caption{(a) Square-root of NOT gate, (b) S gate, (c) Pauli Z gate, (d)
Pauli X gate, (e) Pauli Y gate and (f) Hadamard gate.}
\label{FigPauliOne}
\end{figure}

The braid operator $\mathcal{B}_{1}$ is written in terms of fermion
operators,%
\begin{equation}
\mathcal{B}_{1}=\frac{1}{\sqrt{2}}\left( 1+\gamma _{2}\gamma _{1}\right) =%
\frac{1}{\sqrt{2}}\left( 1+ic_{1}^{\dagger }c_{1}-ic_{1}c_{1}^{\dagger
}\right) ,
\end{equation}%
which operates on two physical qubits (\ref{PsiTwo}) as\cite{IvanovS}%
\begin{equation}
\mathcal{B}_{1}\Psi _{\text{physical}}=e^{-i\pi /4}\left( 
\begin{array}{cccc}
1 & 0 & 0 & 0 \\ 
0 & i & 0 & 0 \\ 
0 & 0 & 1 & 0 \\ 
0 & 0 & 0 & i%
\end{array}%
\right) \left( 
\begin{array}{c}
\left\vert 0,0\right\rangle \\ 
\left\vert 0,1\right\rangle \\ 
\left\vert 1,0\right\rangle \\ 
\left\vert 1,1\right\rangle%
\end{array}%
\right) _{\text{physical}}.
\end{equation}%
Taking the even parity basis, the action is%
\begin{equation}
\mathcal{B}_{1}\Psi _{\text{logical}}=e^{-i\pi /4}\left( 
\begin{array}{cc}
1 & 0 \\ 
0 & i%
\end{array}%
\right) \left( 
\begin{array}{c}
\left\vert 0\right\rangle \\ 
\left\vert 1\right\rangle%
\end{array}%
\right) _{\text{logical}},
\end{equation}%
where the basis for the logical qubit is defined by%
\begin{equation}
\Psi _{\text{logical}}\equiv \left( \left\vert 0\right\rangle
,c_{1}^{\dagger }c_{2}^{\dagger }\left\vert 0\right\rangle \right) ^{t}.
\end{equation}%
The braid operation is written as%
\begin{equation}
\mathcal{B}_{1}=e^{-i\pi /4}U_{\text{S}},  \label{B1US}
\end{equation}%
in terms of the S gate defined by%
\begin{equation}
U_{\text{S}}\equiv \text{diag.}\left( 1,i\right) .  \label{SGate}
\end{equation}%
The braid operator $\mathcal{B}_{2}$ is written in terms of fermion
operators, 
\begin{equation}
\mathcal{B}_{2}=\frac{1}{\sqrt{2}}\left( 1+\gamma _{3}\gamma _{2}\right) =%
\frac{1}{\sqrt{2}}\left( 1+ic_{2}c_{1}^{\dagger }+ic_{2}^{\dagger
}c_{1}^{\dagger }-ic_{2}c_{1}-ic_{2}^{\dagger }c_{1}\right) .
\end{equation}%
It operates on two physical qubits (\ref{PsiTwo}) as\cite{IvanovS},%
\begin{equation}
\mathcal{B}_{2}\Psi _{\text{physical}}=\frac{1}{\sqrt{2}}\left( 
\begin{array}{cccc}
1 & 0 & 0 & -i \\ 
0 & 1 & -i & 0 \\ 
0 & -i & 1 & 0 \\ 
-i & 0 & 0 & 1%
\end{array}%
\right) \left( 
\begin{array}{c}
\left\vert 0,0\right\rangle \\ 
\left\vert 0,1\right\rangle \\ 
\left\vert 1,0\right\rangle \\ 
\left\vert 1,1\right\rangle%
\end{array}%
\right) _{\text{physical}}=U_{xx}\left( 
\begin{array}{c}
\left\vert 0,0\right\rangle \\ 
\left\vert 0,1\right\rangle \\ 
\left\vert 1,0\right\rangle \\ 
\left\vert 1,1\right\rangle%
\end{array}%
\right) _{\text{physical}},
\end{equation}%
where%
\begin{equation}
U_{xx}\equiv \frac{1}{\sqrt{2}}\left( 
\begin{array}{cccc}
1 & 0 & 0 & -i \\ 
0 & 1 & -i & 0 \\ 
0 & -i & 1 & 0 \\ 
-i & 0 & 0 & 1%
\end{array}%
\right) =\exp \left[ -i\frac{\pi }{4}\sigma _{x}\otimes \sigma _{x}\right] .
\label{UMix2}
\end{equation}%
In the even parity basis, the action is%
\begin{equation}
\mathcal{B}_{2}=\frac{1}{\sqrt{2}}\left( 
\begin{array}{cc}
1 & -i \\ 
-i & 1%
\end{array}%
\right) =\exp \left[ -i\frac{\pi }{4}\sigma _{x}\right] \equiv R_{x}.
\label{UMix1}
\end{equation}%
It has the relation%
\begin{equation}
\mathcal{B}_{2}=e^{-i\pi /4}U_{\sqrt{\text{X}}},
\end{equation}%
where $U_{\sqrt{\text{X}}}$ is the square-root of X gate defined by%
\begin{equation}
U_{\sqrt{\text{X}}}\equiv \frac{1}{2}\left( 
\begin{array}{cc}
1+i & 1-i \\ 
1-i & 1+i%
\end{array}%
\right) .
\end{equation}%
The corresponding braiding is shown in Fig.\ref{FigPauliOne}(a).

The braiding operator $\mathcal{B}_{3}$ is written in terms of fermion
operators%
\begin{equation}
\mathcal{B}_{3}=\frac{1}{\sqrt{2}}\left( 1+\gamma _{4}\gamma _{3}\right) =%
\frac{1}{\sqrt{2}}\left( 1+ic_{2}^{\dagger }c_{2}-ic_{2}c_{2}^{\dagger
}\right) ,
\end{equation}%
which operates on two physical qubits (\ref{PsiTwo}) as\cite{IvanovS}%
\begin{equation}
\mathcal{B}_{3}\Psi _{\text{physical}}=e^{-i\pi /4}\left( 
\begin{array}{cccc}
1 & 0 & 0 & 0 \\ 
0 & 1 & 0 & 0 \\ 
0 & 0 & i & 0 \\ 
0 & 0 & 0 & i%
\end{array}%
\right) \left( 
\begin{array}{c}
\left\vert 0,0\right\rangle \\ 
\left\vert 0,1\right\rangle \\ 
\left\vert 1,0\right\rangle \\ 
\left\vert 1,1\right\rangle%
\end{array}%
\right) _{\text{physical}}.
\end{equation}%
In the even parity basis, the action is the same as (\ref{B1US}),%
\begin{equation}
\mathcal{B}_{3}=e^{-i\pi /4}U_{\text{S}},
\end{equation}%
where the S gate is defined by (\ref{SGate}). The corresponding braiding is
shown in Fig.\ref{FigPauliOne}(b).

The Pauli Z gate is given by double braiding of $\mathcal{B}_{3}$,%
\begin{equation}
U_{\text{Z}}\equiv \text{diag.}\left( 1,-1\right) =U_{\text{S}}^{2}=i%
\mathcal{B}_{3}^{2}.
\end{equation}%
The corresponding braiding is shown in Fig.\ref{FigPauliOne}(c).

The Pauli X gate (NOT gate) is given\cite{DasTQCS} by double braiding of $%
\mathcal{B}_{2}$,%
\begin{equation}
U_{\text{X}}\equiv \left( 
\begin{array}{cc}
0 & 1 \\ 
1 & 0%
\end{array}%
\right) =i\mathcal{B}_{2}^{2}.
\end{equation}%
The corresponding braiding is shown in Fig.\ref{FigPauliOne}(d).

Then, the Pauli Y gate is given by sequential applications of $\mathcal{B}%
_{2}$ and $\mathcal{B}_{3}$,%
\begin{equation}
U_{\text{Y}}\equiv \left( 
\begin{array}{cc}
0 & -i \\ 
i & 0%
\end{array}%
\right) =iU_{\text{X}}U_{\text{Z}}=-\mathcal{B}_{2}^{2}\mathcal{B}_{3}^{2}.
\end{equation}%
The corresponding braiding is shown in Fig.\ref{FigPauliOne}(e).

The Hadamard gate is defined by%
\begin{equation}
U_{\text{H}}\equiv \frac{1}{\sqrt{2}}\left( 
\begin{array}{cc}
1 & 1 \\ 
1 & -1%
\end{array}%
\right) .  \label{HadamardGate}
\end{equation}%
It is known to be generated by triple braids as\cite{GeorgievBS,KrausS}%
\begin{equation}
U_{\text{H}}=i\mathcal{B}_{2}\mathcal{B}_{3}\mathcal{B}_{2}.
\end{equation}%
The corresponding braiding is shown in Fig.\ref{FigPauliOne}(f).

\subsection{One Logical qubit entangled states}

The even cat state is made by applying the Hadamard gate (\ref{HadamardGate}%
) as%
\begin{equation}
U_{\text{H}}\left\vert 0\right\rangle _{\text{logical}}=i\mathcal{B}_{2}%
\mathcal{B}_{3}\mathcal{B}_{2}\left\vert 0\right\rangle _{\text{logical}}=%
\frac{1}{\sqrt{2}}\left( \left\vert 0\right\rangle _{\text{logical}%
}+\left\vert 1\right\rangle _{\text{logical}}\right) .
\end{equation}%
However, a double braiding is enough for the construction of the even cat
state $\left\vert \psi \right\rangle _{\text{even-cat}}$,%
\begin{equation}
e_{1}^{i\pi /4}\mathcal{B}_{1}\mathcal{B}_{2}\left\vert 0\right\rangle _{%
\text{logical}}=\frac{1}{\sqrt{2}}\left( \left\vert 0\right\rangle _{\text{%
logical}}+\left\vert 1\right\rangle _{\text{logical}}\right) \equiv
\left\vert \psi \right\rangle _{\text{even-cat}}.
\end{equation}%
On the other hand, the odd cat state $\left\vert \psi \right\rangle _{\text{%
odd-cat}}$ is made as%
\begin{equation}
e^{-i\pi /4}\mathcal{B}_{1}^{-1}\mathcal{B}_{2}\left\vert 0\right\rangle _{%
\text{logical}}=\frac{1}{\sqrt{2}}\left( \left\vert 0\right\rangle _{\text{%
logical}}-\left\vert 1\right\rangle _{\text{logical}}\right) \equiv
\left\vert \psi \right\rangle _{\text{odd-cat}}.
\end{equation}

Only 6 states can be constructed by braiding in one qubit. The state $\left(
\left\vert 0\right\rangle _{\text{logical}}\pm i\left\vert 1\right\rangle _{%
\text{logical}}\right) /\sqrt{2}$ is constructed by single braiding, while
the states $\left\vert 1\right\rangle _{\text{logical}} $ and $\left(
\left\vert 0\right\rangle _{\text{logical}}\pm \left\vert 1\right\rangle _{%
\text{logical}}\right) /\sqrt{2}$ are constructed by double braiding. No
further states can be constructed by further braiding.

\subsection{Two logical qubits}

In order to construct two logical qubits, we use six Majorana fermions $%
\gamma _{1}$, $\gamma _{2}$, $\gamma _{3}$, $\gamma _{4}$, $\gamma _{5}$ and 
$\gamma _{6}$. Three ordinary fermion operators are given by 
\begin{equation}
c_{1}=\frac{1}{2}\left( \gamma _{1}+i\gamma _{2}\right) ,\quad c_{2}=\frac{1%
}{2}\left( \gamma _{3}+i\gamma _{4}\right) ,\quad c_{3}=\frac{1}{2}\left(
\gamma _{5}+i\gamma _{6}\right) .
\end{equation}%
The basis of phycial qubits are given by%
\begin{align}
\Psi _{\text{physical}}& =(\left\vert 0\right\rangle ,c_{1}^{\dagger
}\left\vert 0\right\rangle ,c_{2}^{\dagger }\left\vert 0\right\rangle
,c_{1}^{\dagger }c_{2}^{\dagger }\left\vert 0\right\rangle ,c_{3}^{\dagger
}\left\vert 0\right\rangle ,c_{1}^{\dagger }c_{3}^{\dagger }\left\vert
0\right\rangle ,c_{2}^{\dagger }c_{3}^{\dagger }\left\vert 0\right\rangle
,c_{1}^{\dagger }c_{2}^{\dagger }c_{3}^{\dagger }\left\vert 0\right\rangle
)^{t}  \notag \\
& \equiv (\left\vert 0,0,0\right\rangle _{\text{physical}},\left\vert
0,0,1\right\rangle _{\text{physical}},\left\vert 0,1,0\right\rangle _{\text{%
physical}},\left\vert 0,1,1\right\rangle _{\text{physical}},\left\vert
1,0,0\right\rangle _{\text{physical}},  \notag \\
& \quad \quad \quad \quad \quad \quad \left\vert 1,0,1\right\rangle _{\text{%
physical}},\left\vert 1,1,0\right\rangle _{\text{physical}},\left\vert
1,1,1\right\rangle _{\text{physical}})^{t}.
\end{align}

The explicit braid operators on the physical qubits are%
\begin{align}
\mathcal{B}_{1}& =I_{2}\otimes I_{2}\otimes U_{\text{S}},  \notag \\
\mathcal{B}_{2}& =I_{2}\otimes U_{xx},  \notag \\
\mathcal{B}_{3}& =I_{2}\otimes U_{\text{S}}\otimes I_{2},  \notag \\
\mathcal{B}_{4}& =U_{xx}\otimes I_{2},  \notag \\
\mathcal{B}_{5}& =U_{\text{S}}\otimes I_{2}\otimes I_{2}.
\end{align}%
Two logical qubits are constructed from three physical qubits as%
\begin{equation}
\left( 
\begin{array}{c}
\left\vert 0,0\right\rangle \\ 
\left\vert 0,1\right\rangle \\ 
\left\vert 1,0\right\rangle \\ 
\left\vert 1,1\right\rangle%
\end{array}%
\right) _{\text{logical}}=\left( 
\begin{array}{c}
\left\vert 0,0,0\right\rangle \\ 
\left\vert 0,1,1\right\rangle \\ 
\left\vert 1,0,1\right\rangle \\ 
\left\vert 1,1,0\right\rangle%
\end{array}%
\right) _{\text{physical}}.
\end{equation}%
In the logical qubit basis, the braiding operators are%
\begin{align}
\mathcal{B}_{1}& =e^{-i\pi /4}\text{diag.}\left( 1,i,i,1\right) ,  \notag \\
\mathcal{B}_{2}& =I_{2}\otimes R_{x},  \notag \\
\mathcal{B}_{3}& =e^{-i\pi /4}\text{diag.}\left( 1,i,1,i\right) ,  \notag \\
\mathcal{B}_{4}& =U_{xx},  \notag \\
\mathcal{B}_{5}& =e^{-i\pi /4}\text{diag.}\left( 1,1,i,i\right) .
\end{align}%
where $R_{x}$ is defined by (\ref{UMix1}) and $U_{xx}$ is defined by (\ref%
{UMix2}).

\begin{figure}[t]
\centerline{\includegraphics[width=0.88\textwidth]{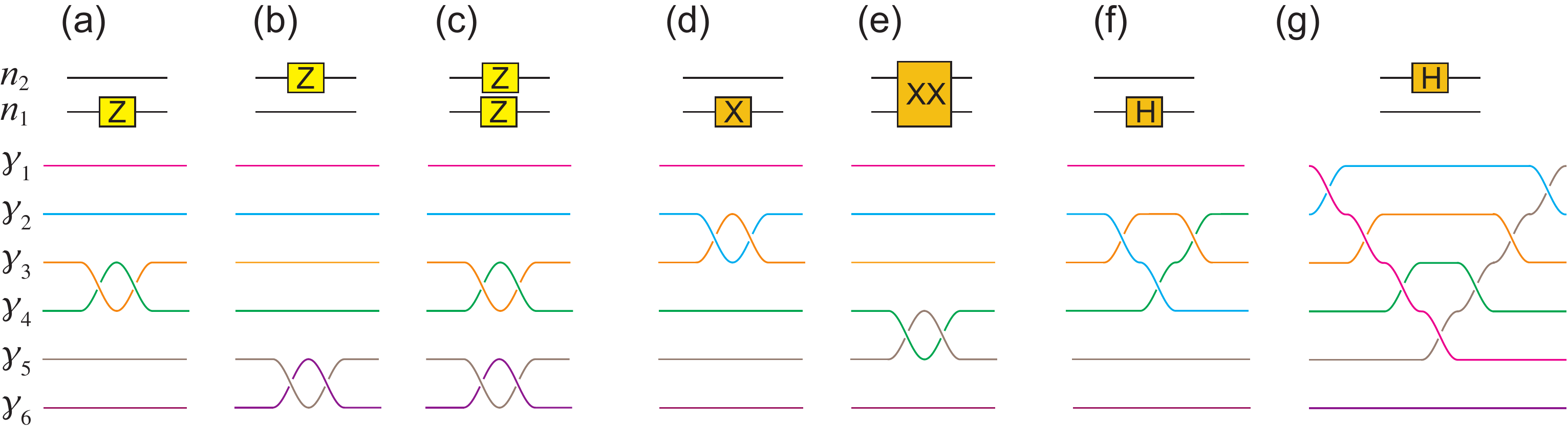}}
\caption{The braiding process for Pauli gates. (a) Pauli Z gate embedded to
the first qubit, (b) Pauli Z gate embedded to the second qubit, (c) Two
Pauli Z gates are embedded to the first and the second qubits, (d) Pauli X
gate embedded to the first qubit, (e) Two Pauli X gates are embedded to the
first and the second qubits, (f) Hadamard gate embedded to the first qubit
and (g) Hadamard gate embedded to the second qubit. }
\label{FigPauliTwo}
\end{figure}

\subsection{Pauli gates}

The two-qubit Pauli gates are defined by%
\begin{equation}
\sigma _{k_{2}}\otimes \sigma _{k_{1}},
\end{equation}%
where $k_{1}$ and $k_{2}$ take $0,x,y$ and $z$. The Pauli Z gates are
generated by braiding $\mathcal{B}_{2k+1}$ with odd indices,%
\begin{equation}
I_{2}\otimes \sigma _{\text{Z}}=i\mathcal{B}_{3}^{2},\quad \sigma _{\text{Z}%
}\otimes I_{2}=i\mathcal{B}_{5}^{2},\quad \sigma _{\text{Z}}\otimes \sigma _{%
\text{Z}}=-\mathcal{B}_{5}^{2}\mathcal{B}_{3}^{2}.
\end{equation}%
They are summarized as%
\begin{equation}
\left( \sigma _{\text{Z}}\right) ^{n_{2}}\otimes \left( \sigma _{\text{Z}%
}\right) ^{n_{1}}=\left( i\mathcal{B}_{5}^{2}\right) ^{n_{2}}\left( i%
\mathcal{B}_{3}^{2}\right) ^{n_{1}},
\end{equation}%
where $n_{1}$ and $n_{2}$ take $0$ or $1$.

The Pauli X gates are generated by braiding with even indices $\mathcal{B}%
_{2k}$,%
\begin{equation}
I_{2}\otimes \sigma _{\text{X}}=i\mathcal{B}_{2}^{2},\quad \sigma _{\text{X}%
}\otimes \sigma _{\text{X}}=i\mathcal{B}_{4}^{2},\quad I_{2}\otimes \sigma _{%
\text{X}}=-\mathcal{B}_{4}^{2}\mathcal{B}_{2}^{2}.
\end{equation}%
It should be noted that $\mathcal{B}_{4}^{2}$ does not generate $%
I_{2}\otimes \sigma _{\text{X}}$ but generate $\sigma _{\text{X}}\otimes
\sigma _{\text{X}}$. We show the braiding for Pauli gates in Fig.\ref%
{FigPauliTwo}.

Pauli Y gates are generated by sequential applications of Pauli X gates and
Pauli Z gates based on the relation $U_{\text{Y}}=iU_{\text{X}}U_{\text{Z}}$%
. Thus, all of Pauli gates for two qubits can be generated by braiding.

\subsection{Hadamard gates}

The Hadamard gate acting on the first qubit can be embedded as%
\begin{equation}
I_{2}\otimes U_{\text{H}}=i\mathcal{B}_{2}\mathcal{B}_{3}\mathcal{B}_{2}.
\label{B232}
\end{equation}%
The Hadamard gate acting on the second qubit can be embedded as%
\begin{equation}
U_{\text{H}}\otimes I_{2}=-\mathcal{B}_{1}\mathcal{B}_{2}\mathcal{B}_{3}%
\mathcal{B}_{4}\mathcal{B}_{3}\mathcal{B}_{2}\mathcal{B}_{1}.
\end{equation}%
It requires more braiding than the previous results\cite{GeorgievNPBS,KrausS}%
, where three braiding are enough. It is due to the choice of the
correspondence between the physical and logical qubits.

\begin{figure}[t]
\centerline{\includegraphics[width=0.88\textwidth]{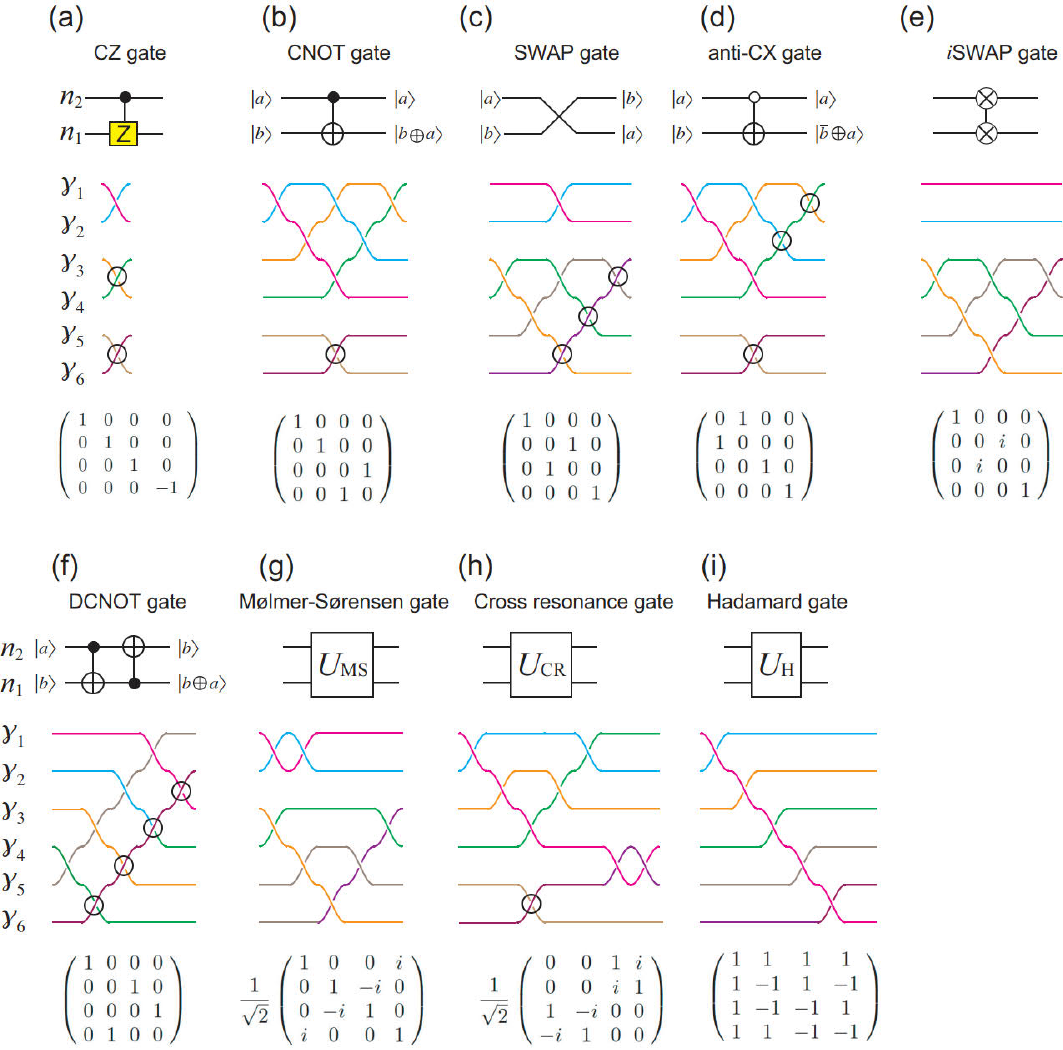}}
\caption{Braiding process for various two-qubit quantum gates. (a) CZ gate,
(b) CNOT gate, (c) SWAP gate, (d) anti-CX gate, (e) $i$SWAP gate, (f) DCNOT
gate, (g) Molmer-Sorensen gate, (h) cross-resonance gate (i) Hadamard gate.}
\label{FigCNOT}
\end{figure}

\subsection{Quantum gates for two logical qubits}

It is known that the controlled-Z (CZ) gate 
\begin{equation}
U_{\text{CZ}}=\text{diag.}\left( 1,1,1,-1\right)
\end{equation}%
is generated as\cite{GeorgievNPBS}%
\begin{equation}
U_{\text{CZ}}=e^{-i\pi /4}\mathcal{B}_{5}^{-1}\left( \mathcal{B}_{3}\right)
^{-1}\mathcal{B}_{1}.
\end{equation}%
See Fig.\ref{FigCNOT}(a).

It is also known that the controlled-NOT (CNOT) gate%
\begin{equation}
U_{\text{CNOT}}=\left( 
\begin{array}{cccc}
1 & 0 & 0 & 0 \\ 
0 & 1 & 0 & 0 \\ 
0 & 0 & 0 & 1 \\ 
0 & 0 & 1 & 0%
\end{array}%
\right)
\end{equation}%
is generated by 7 braiding\cite{GeorgievBS,GeorgievNPBS,AhlS}, wher braiding
are given by%
\begin{equation}
U_{\text{CNOT}}=-e^{-i\pi /4}\mathcal{B}_{5}^{-1}\mathcal{B}_{1}\mathcal{B}%
_{2}\mathcal{B}_{3}\mathcal{B}_{1}\mathcal{B}_{2}\mathcal{B}_{1}.
\end{equation}%
See Fig.\ref{FigCNOT}(b). On the other hand, there is a quantum circuit
decomposition formula%
\begin{equation}
U_{\text{CNOT}}=\left( I_{2}\otimes U_{\text{H}}\right) U_{\text{CZ}}\left(
I_{2}\otimes U_{\text{H}}\right) ,
\end{equation}%
which involves 9 braiding.

The SWAP gate is defined by 
\begin{equation}
U_{\text{SWAP}}\equiv \left( 
\begin{array}{cccc}
1 & 0 & 0 & 0 \\ 
0 & 0 & 1 & 0 \\ 
0 & 1 & 0 & 0 \\ 
0 & 0 & 0 & 1%
\end{array}%
\right) ,
\end{equation}%
which is realized by 7 braiding as%
\begin{equation}
U_{\text{SWAP}}=e^{i\pi /4}\left( \mathcal{B}_{3}\right) ^{-1}\left( 
\mathcal{B}_{4}\right) ^{-1}\left( \mathcal{B}_{5}\right) ^{-1}\mathcal{B}%
_{3}\mathcal{B}_{4}\mathcal{B}_{3}\mathcal{B}_{1}.
\end{equation}%
See Fig.\ref{FigCNOT}(c). This is smaller than the previous result using 15
braiding\cite{GeorgievNPBS} based on the quantum circuit decomposition%
\begin{equation}
U_{\text{SWAP}}=\left( I_{2}\otimes U_{\text{H}}\right) U_{\text{CZ}}\left(
I_{2}\otimes U_{\text{H}}\right) \left( U_{\text{H}}\otimes I_{2}\right) U_{%
\text{CZ}}\left( U_{\text{H}}\otimes I_{2}\right) \left( I_{2}\otimes U_{%
\text{H}}\right) U_{\text{CZ}}\left( I_{2}\otimes U_{\text{H}}\right) .
\end{equation}

We list up various quantum gates generated by braiding.

The anti-CNOT gate is defined by\cite{WilliamsS}%
\begin{equation}
U_{\text{\={C}X}}\equiv \left( 
\begin{array}{cccc}
0 & 1 & 0 & 0 \\ 
1 & 0 & 0 & 0 \\ 
0 & 0 & 1 & 0 \\ 
0 & 0 & 0 & 1%
\end{array}%
\right) ,
\end{equation}%
which is generated by 7 braiding%
\begin{equation}
U_{\text{\={C}X}}=e^{i\pi /4}\mathcal{B}_{5}^{-1}\mathcal{B}_{1}^{-1}%
\mathcal{B}_{2}^{-1}\mathcal{B}_{3}\mathcal{B}_{1}\mathcal{B}_{2}\mathcal{B}%
_{1}.
\end{equation}%
It can be decomposed into $U_{\text{\={C}X}}=\left( I_{2}\otimes U_{\text{X}%
}\right) U_{\text{CNOT}}$. If we use this relation, 9 braiding are
necessary. See Fig.\ref{FigCNOT}(d).

The $i$SWAP gate is defined by%
\begin{equation}
U_{i\text{SWAP}}\equiv \left( 
\begin{array}{cccc}
1 & 0 & 0 & 0 \\ 
0 & 0 & i & 0 \\ 
0 & i & 0 & 0 \\ 
0 & 0 & 0 & 1%
\end{array}%
\right) ,
\end{equation}%
which is realized by the six braiding%
\begin{equation}
U_{i\text{SWAP}}=-\mathcal{B}_{3}\mathcal{B}_{4}\mathcal{B}_{5}\mathcal{B}%
_{3}\mathcal{B}_{4}\mathcal{B}_{3}.
\end{equation}%
See Fig.\ref{FigCNOT}(e).

The double CNOT gate is defined by\cite{CollinsS}%
\begin{equation}
U_{\text{DCNOT}}\equiv \left( 
\begin{array}{cccc}
1 & 0 & 0 & 0 \\ 
0 & 0 & 1 & 0 \\ 
0 & 0 & 0 & 1 \\ 
0 & 1 & 0 & 0%
\end{array}%
\right) ,
\end{equation}%
which is realized by%
\begin{equation}
U_{\text{DCNOT}}=\mathcal{B}_{2}^{-1}\mathcal{B}_{3}^{-1}\mathcal{B}_{4}^{-1}%
\mathcal{B}_{5}^{-1}\mathcal{B}_{1}\mathcal{B}_{2}\mathcal{B}_{3}\mathcal{B}%
_{4}.
\end{equation}%
See Fig.\ref{FigCNOT}(f).

The M\o lmer-S\o rensen gate is defined by\cite{MSS}%
\begin{equation}
U_{\text{MS}}\equiv \frac{1}{\sqrt{2}}\left( 
\begin{array}{cccc}
1 & 0 & 0 & i \\ 
0 & 1 & -i & 0 \\ 
0 & -i & 1 & 0 \\ 
i & 0 & 0 & 1%
\end{array}%
\right) ,
\end{equation}%
which is realized by%
\begin{equation}
U_{\text{MS}}=-i\mathcal{B}_{3}\mathcal{B}_{4}\mathcal{B}_{5}\mathcal{B}_{4}%
\mathcal{B}_{3}\mathcal{B}_{1}\mathcal{B}_{1}.
\end{equation}%
See Fig.\ref{FigCNOT}(g).

The cross-resonance gate is defined by\cite{CorcoS}%
\begin{equation}
U_{\text{CR}}\equiv \frac{1}{\sqrt{2}}\left( 
\begin{array}{cccc}
0 & 0 & 1 & i \\ 
0 & 0 & i & 1 \\ 
1 & -i & 0 & 0 \\ 
-i & 1 & 0 & 0%
\end{array}%
\right) ,
\end{equation}%
which is realized by%
\begin{equation}
U_{\text{CR}}=-\mathcal{B}_{4}\mathcal{B}_{4}\mathcal{B}_{1}\mathcal{B}_{2}%
\mathcal{B}_{3}\mathcal{B}_{2}\mathcal{B}_{1}.
\end{equation}%
See Fig.\ref{FigCNOT}(h).

We define the entangled Hadamard gate by%
\begin{equation}
U_{\text{H}}^{\left( 2\right) }=\left( 
\begin{array}{cccc}
1 & 1 & 1 & 1 \\ 
1 & -1 & 1 & -1 \\ 
1 & -1 & -1 & 1 \\ 
1 & 1 & -1 & -1%
\end{array}%
\right) ,
\end{equation}%
which is realized by%
\begin{equation}
U_{\text{H}}^{\left( 2\right) }=-e^{-i\pi /4}\mathcal{B}_{5}\mathcal{B}_{4}%
\mathcal{B}_{3}\mathcal{B}_{2}\mathcal{B}_{1}.
\end{equation}%
See Fig.\ref{FigCNOT}(i). It is different from the cross product of the
Hadamard gates%
\begin{equation}
U_{\text{H}}^{\left( 2\right) }\neq U_{\text{H}}\otimes U_{\text{H}}.
\end{equation}%
We note that it is obtained by a permutation of the third and fourth columns
of the cross product of the Hadamard gates given by%
\begin{equation}
U_{\text{H}}\otimes U_{\text{H}}=\left( 
\begin{array}{cccc}
1 & 1 & 1 & 1 \\ 
1 & -1 & 1 & -1 \\ 
1 & 1 & -1 & -1 \\ 
1 & -1 & -1 & 1%
\end{array}%
\right) ,
\end{equation}%
which leads to a relation%
\begin{equation}
U_{\text{H}}\otimes U_{\text{H}}=U_{\text{CNOT}}U_{\text{H}}^{\left(
2\right) }.
\end{equation}%
Hence, it is realized by%
\begin{equation}
U_{\text{H}}\otimes U_{\text{H}}=-\mathcal{B}_{5}^{-1}\mathcal{B}_{1}%
\mathcal{B}_{2}\mathcal{B}_{3}\mathcal{B}_{1}\mathcal{B}_{2}\mathcal{B}_{1}%
\mathcal{B}_{5}\mathcal{B}_{4}\mathcal{B}_{3}\mathcal{B}_{2}\mathcal{B}_{1}.
\end{equation}%
Both $U_{\text{H}}^{\left( 2\right) }$ and $U_{\text{H}}\otimes U_{\text{H}}$%
\ are the Hadamard gates and they are useful for various quantum algorithms.

\subsection{Equal-coefficient states}

The equal-coefficient state is constructed as%
\begin{align}
i\mathcal{B}_{1}\mathcal{B}_{2}\mathcal{B}_{3}\mathcal{B}_{4}\mathcal{B}%
_{5}\left\vert 0,0\right\rangle _{\text{logical}}& =\frac{1}{2}\left(
\left\vert 0,0\right\rangle _{\text{logical}}+\left\vert 0,1\right\rangle _{%
\text{logical}}+\left\vert 1,0\right\rangle _{\text{logical}}+\left\vert
1,1\right\rangle _{\text{logical}}\right)  \notag \\
& \equiv \frac{1}{2}\left( \left\vert 0\right\rangle _{\text{logical}}^{%
\text{decimal}}+\left\vert 1\right\rangle _{\text{logical}}^{\text{decimal}%
}+\left\vert 2\right\rangle _{\text{logical}}^{\text{decimal}}+\left\vert
3\right\rangle _{\text{logical}}^{\text{decimal}}\right) ,
\end{align}%
where $\left\vert j\right\rangle _{\text{logical}}^{\text{decimal}}$ is a
decimal representation of qubits. It is a fundamental entangled state for
two qubits.

\subsection{Three logical qubits}

We use eight Majorana fermions in order to construct three logical qubits,%
\begin{equation}
c_{1}=\frac{1}{2}\left( \gamma _{1}+i\gamma _{2}\right) ,\quad c_{2}=\frac{1%
}{2}\left( \gamma _{3}+i\gamma _{4}\right) ,\quad c_{3}=\frac{1}{2}\left(
\gamma _{5}+i\gamma _{6}\right) .\quad c_{4}=\frac{1}{2}\left( \gamma
_{7}+i\gamma _{8}\right) .
\end{equation}%
The explicit braid actions on the physical qubits are%
\begin{align}
\mathcal{B}_{1}& =I_{2}\otimes I_{2}\otimes I_{2}\otimes U_{\text{S}}, 
\notag \\
\mathcal{B}_{2}& =I_{2}\otimes I_{2}\otimes U_{xx},  \notag \\
\mathcal{B}_{3}& =I_{2}\otimes I_{2}\otimes U_{\text{S}}\otimes I_{2}, 
\notag \\
\mathcal{B}_{4}& =I_{2}\otimes U_{xx}\otimes I_{2},  \notag \\
\mathcal{B}_{5}& =I_{2}\otimes U_{\text{S}}\otimes I_{2}\otimes I_{2}, 
\notag \\
\mathcal{B}_{6}& =U_{xx}\otimes I_{2}\otimes I_{2},  \notag \\
\mathcal{B}_{7}& =U_{\text{S}}\otimes I_{2}\otimes I_{2}\otimes I_{2}.
\end{align}%
Three logical qubits are constructed from four physical qubits as%
\begin{equation}
\left( 
\begin{array}{c}
\left\vert 0,0,0\right\rangle  \\ 
\left\vert 0,0,1\right\rangle  \\ 
\left\vert 0,1,0\right\rangle  \\ 
\left\vert 0,1,1\right\rangle  \\ 
\left\vert 1,0,0\right\rangle  \\ 
\left\vert 1,0,1\right\rangle  \\ 
\left\vert 1,1,0\right\rangle  \\ 
\left\vert 1,1,1\right\rangle 
\end{array}%
\right) _{\text{logical}}=\left( 
\begin{array}{c}
\left\vert 0,0,0,0\right\rangle  \\ 
\left\vert 0,0,1,1\right\rangle  \\ 
\left\vert 0,1,0,1\right\rangle  \\ 
\left\vert 0,1,1,0\right\rangle  \\ 
\left\vert 1,0,0,1\right\rangle  \\ 
\left\vert 1,0,1,0\right\rangle  \\ 
\left\vert 1,1,0,0\right\rangle  \\ 
\left\vert 1,1,1,1\right\rangle 
\end{array}%
\right) _{\text{physical}}.
\end{equation}%
Explicit matrix representations for the braiding operator are%
\begin{align}
\mathcal{B}_{1}& =e^{-i\pi /4}\text{diag.}\left( 1,i,i,1,i,1,1,i\right)
=\exp \left[ -\frac{i\pi }{4}\sigma _{z}\otimes \sigma _{z}\otimes \sigma
_{z}\right] ,  \notag \\
\mathcal{B}_{2}& =I_{4}\otimes R_{x},  \notag \\
\mathcal{B}_{3}& =e^{-i\pi /4}\text{diag.}\left( 1,i,1,i,1,i,1,i\right)
=e^{-i\pi /4}I_{4}\otimes U_{\text{S}},  \notag \\
\mathcal{B}_{4}& =I_{2}\otimes U_{xx},  \notag \\
\mathcal{B}_{5}& =e^{-i\pi /4}\text{diag.}\left( 1,1,i,i,1,1,i,i\right)
=e^{-i\pi /4}I_{2}\otimes U_{\text{S}}\otimes I_{2},  \notag \\
\mathcal{B}_{6}& =U_{xx}\otimes I_{2},  \notag \\
\mathcal{B}_{7}& =e^{-i\pi /4}\text{diag.}\left( 1,1,1,1,i,i,i,i\right)
=e^{-i\pi /4}U_{\text{S}}\otimes I_{4}.
\end{align}

\subsection{Pauli gates}

The three-qubit Pauli gates are defined by%
\begin{equation}
\sigma _{k_{3}}\otimes \sigma _{k_{2}}\otimes \sigma _{k_{1}},
\end{equation}%
where $k_{1}$, $k_{2}$ and $k_{3}$ take $0,x,y$ and $z$. The Pauli Z gates
are generated by braiding operators $\mathcal{B}_{2k+1}$ with odd indices%
\begin{equation}
I_{2}\otimes I_{2}\otimes \sigma _{\text{Z}}=i\mathcal{B}_{3}^{2},\quad
I_{2}\otimes \sigma _{\text{Z}}\otimes I_{2}=i\mathcal{B}_{5}^{2},\quad
\sigma _{\text{Z}}\otimes I_{2}\otimes I_{2}=i\mathcal{B}_{7}^{2},
\end{equation}%
They are summarized as%
\begin{equation}
\left( \sigma _{\text{Z}}\right) ^{n_{3}}\otimes \left( \sigma _{\text{Z}%
}\right) ^{n_{2}}\otimes \left( \sigma _{\text{Z}}\right) ^{n_{1}}=\left( i%
\mathcal{B}_{7}^{2}\right) ^{n_{3}}\left( i\mathcal{B}_{5}^{2}\right)
^{n_{2}}\left( i\mathcal{B}_{3}^{2}\right) ^{n_{1}},
\end{equation}%
where $n_{1}$ $n_{2}$ and $n_{3}$ take $0$ or $1$.

\begin{figure}[t]
\centerline{\includegraphics[width=0.68\textwidth]{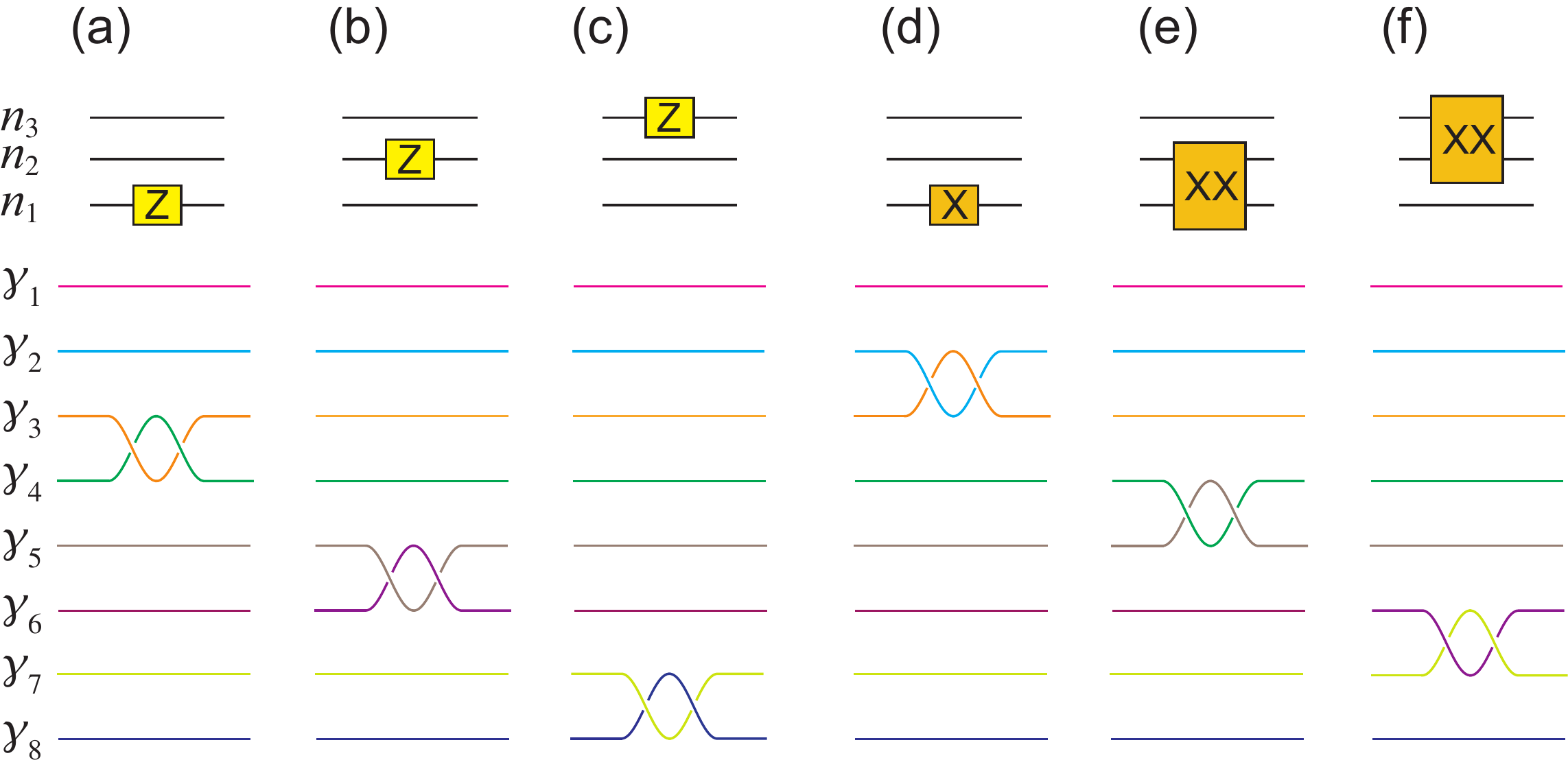}}
\caption{Pauli gates embedded in three qubits. (a) Pauli Z gate embedded to
the first qubit, (b) Pauli Z gate embedded to the second qubit, (c) Pauli Z
gate embedded to the third qubit, (d) Pauli X gate embedded to the first
qubit, (e) Two Pauli X gates are embedded to the first and second qubits and
(f) Two Pauli X gates are embedded to the second third qubits.}
\label{FigPauliThree}
\end{figure}

The Pauli X gates are generated by braiding operators with even numbers,%
\begin{equation}
I_{2}\otimes I_{2}\otimes \sigma _{\text{X}}=i\mathcal{B}_{2}^{2},\quad
I_{2}\otimes \sigma _{\text{X}}\otimes \sigma _{\text{X}}=i\mathcal{B}%
_{4}^{2},\quad \sigma _{\text{X}}\otimes \sigma _{\text{X}}\otimes I_{2}=i%
\mathcal{B}_{6}^{2}.
\end{equation}%
We show the corresponding braiding in Fig.\ref{FigPauliThree}. It is
impossible to construct logical gates corresponding to%
\begin{equation}
I_{2}\otimes \sigma _{\text{X}}\otimes I_{2}\text{ and }\sigma _{\text{X}%
}\otimes I_{2}\otimes I_{2}
\end{equation}%
solely by braiding. This problem is solved by introducing many-body
interactions of Majorana fermions as in Eq.(\ref{table}).

The other Pauli gates can be generated by sequential applications of the
above Pauli gates.

\subsection{Diagonal braiding}

We first search braiding operators for the quantum gates generated by odd
double braiding, 
\begin{equation}
U_{\text{diag}}=\left( \mathcal{B}_{7}^{2}\right) ^{n_{3}}\left( \mathcal{B}%
_{5}^{2}\right) ^{n_{2}}\left( \mathcal{B}_{3}^{2}\right) ^{n_{1}}.
\end{equation}%
There are eight patterns represented by the Pauli Z gates%
\begin{align}
\text{diag.}\left( 1,1,1,1,1,1,1,1\right) & =I_{2}\otimes I_{2}\otimes I_{2},
\notag \\
\text{diag.}\left( 1,-1,1,-1,1,-1,1,-1\right) & =I_{2}\otimes I_{2}\otimes
\sigma _{\text{Z}}=i\mathcal{B}_{3}^{2},  \notag \\
\text{diag.}\left( 1,1,-1,-1,1,1,-1,-1\right) & =I_{2}\otimes \sigma _{\text{%
Z}}\otimes I_{2}=i\mathcal{B}_{5}^{2},  \notag \\
\text{diag.}\left( 1,1,1,1,-1,-1,-1,-1\right) & =\sigma _{\text{Z}}\otimes
I_{2}\otimes I_{2}=i\mathcal{B}_{7}^{2},  \notag \\
\text{diag.}\left( 1,-1,-1,1,1,-1,-1,1\right) & =I_{2}\otimes \sigma _{\text{%
Z}}\otimes \sigma _{\text{Z}}=-\mathcal{B}_{5}^{2}\mathcal{B}_{3}^{2}, 
\notag \\
\text{diag.}\left( 1,1,-1,-1,-1,-1,1,1\right) & =\sigma _{\text{Z}}\otimes
\sigma _{\text{Z}}\otimes I_{2}=-\mathcal{B}_{7}^{2}\mathcal{B}_{5}^{2}, 
\notag \\
\text{diag.}\left( 1,-1,1,-1,-1,1,-1,1\right) & =\sigma _{\text{Z}}\otimes
I_{2}\otimes \sigma _{\text{Z}}=-\mathcal{B}_{7}^{2}\mathcal{B}_{3}^{2}, 
\notag \\
\text{diag.}\left( 1,-1,-1,1,-1,1,1,-1\right) & =\sigma _{\text{Z}}\otimes
\sigma _{\text{Z}}\otimes \sigma _{\text{Z}}=-i\mathcal{B}_{7}^{2}\mathcal{B}%
_{5}^{2}\mathcal{B}_{3}^{2}.
\end{align}

Next, we search real and diagonal gates obtained by the following odd
braiding%
\begin{equation}
\left( \mathcal{B}_{7}\right) ^{n_{3}}\left( \mathcal{B}_{5}\right)
^{n_{2}}\left( \mathcal{B}_{3}\right) ^{n_{1}}.
\end{equation}%
We search states whose components are $\pm 1$. There are four additional
quantum gates, whose traces are zero Tr$U_{\text{diag}}=0$,%
\begin{align}
& \text{diag.}\left( 1,-1,-1,-1,1,1,1,-1\right) =i\mathcal{B}_{4}^{-1}%
\mathcal{B}_{3}\mathcal{B}_{2}\mathcal{B}_{1},  \notag \\
& \text{diag.}\left( 1,-1,1,1,-1,-1,1,-1\right) =i\mathcal{B}_{3}^{-1}%
\mathcal{B}_{4}\mathcal{B}_{2}\mathcal{B}_{1},  \notag \\
& \text{diag.}\left( 1,1,-1,1,-1,1,-1,-1\right) =i\mathcal{B}_{2}^{-1}%
\mathcal{B}_{3}\mathcal{B}_{4}\mathcal{B}_{1},  \notag \\
& \text{diag.}\left( 1,1,1,-1,1,-1,-1,-1\right) =-i\mathcal{B}_{4}^{-1}%
\mathcal{B}_{3}^{-1}\mathcal{B}_{2}^{-1}\mathcal{B}_{1}.
\end{align}%
In addition, there are additional quantum gates, whose traces are nonzero Tr$%
U_{\text{diag}}\neq 0$,%
\begin{align}
& \text{diag.}\left( 1,-1,-1,-1,-1,-1,-1,1\right) =-\mathcal{B}_{4}\mathcal{B%
}_{3}\mathcal{B}_{2}\mathcal{B}_{1},  \notag \\
& \text{diag.}\left( 1,-1,1,1,1,1,-1,1\right) =\mathcal{B}_{4}^{-1}\mathcal{B%
}_{3}^{-1}\mathcal{B}_{2}\mathcal{B}_{1},  \notag \\
& \text{diag.}\left( 1,1,-1,1,1,-1,1,1\right) =\mathcal{B}_{4}^{-1}\mathcal{B%
}_{2}^{-1}\mathcal{B}_{3}\mathcal{B}_{1},  \notag \\
& \text{diag.}\left( 1,1,1,-1,-1,1,1,1\right) =\mathcal{B}_{3}^{-1}\mathcal{B%
}_{2}^{-1}\mathcal{B}_{4}\mathcal{B}_{1}.
\end{align}

It is natural to anticipate that the CZ gate and the CCZ gate are generated
by even braiding because they are diagonal gates. However, this is not the
case by checking all $4^{3}$ patterns of braiding. As a result, the even
braiding do not generate the CZ gates%
\begin{align}
I_{2}\otimes U_{\text{CZ}}& =\text{diag.}\left( 1,1,1,-1,1,1,1,-1\right) , 
\notag \\
U_{\text{CZ}}\otimes I_{2}& =\text{diag.}\left( 1,1,1,1,1,1,-1,-1\right) ,
\end{align}%
and the CCZ gate 
\begin{equation}
U_{\text{CCZ}}=\text{diag.}\left( 1,1,1,1,1,1,1,-1\right) .
\end{equation}%
This problem is solved by introducing many-body interactions of Majorana
fermions as shown in the main text.

\subsection{Hadamard gates}

The Hadamard gate can be embedded in the first qubit as%
\begin{equation}
I_{2}\otimes I_{2}\otimes U_{\text{H}}=i\mathcal{B}_{2}\mathcal{B}_{3}%
\mathcal{B}_{2},
\end{equation}%
as in the case of (\ref{B232}). We also find that the Hadamard gate can be
embedded in the third qubit as%
\begin{equation}
U_{\text{H}}\otimes I_{2}\otimes I_{2}=-i\mathcal{B}_{1}\mathcal{B}_{2}%
\mathcal{B}_{3}\mathcal{B}_{4}\mathcal{B}_{5}\mathcal{B}_{6}\mathcal{B}_{5}%
\mathcal{B}_{4}\mathcal{B}_{3}\mathcal{B}_{2}\mathcal{B}_{1}.
\end{equation}%
On the other hand, it is very hard to embed the Hadamard gate in the second
qubit $I_{2}\otimes U_{\text{H}}\otimes I_{2}$.\ It is possible by
introducing many-body interactions of Majorana fermions. The Hadamard gate
for the $N$-th qubit is given by%
\begin{equation}
U_{\text{H}}\otimes I_{2N-2}\propto \mathcal{B}_{1}\mathcal{B}_{2}\cdots 
\mathcal{B}_{2N-1}\mathcal{B}_{2N}\mathcal{B}_{2N-1}\cdots \mathcal{B}_{2}%
\mathcal{B}_{1}.
\end{equation}

\subsection{Two-qubit quantum gates embedded in three-qubit quantum gates}

The $i$SWAP gate can be embedded to a three-qubit topological gate because
it does not involve $\mathcal{B}_{1}$ and is given by%
\begin{equation}
I_{2}\otimes U_{i\text{SWAP}}=-\mathcal{B}_{3}\mathcal{B}_{4}\mathcal{B}_{5}%
\mathcal{B}_{3}\mathcal{B}_{4}\mathcal{B}_{3}.
\end{equation}%
See Fig.\ref{FourBitH}(a). We also find the $i$SWAP gate can be embedded as%
\begin{equation}
U_{i\text{SWAP}}\otimes I_{2}=-\mathcal{B}_{5}\mathcal{B}_{6}\mathcal{B}_{7}%
\mathcal{B}_{5}\mathcal{B}_{6}\mathcal{B}_{5}.
\end{equation}%
See Fig.\ref{FourBitH}(b).

\subsection{Three-qubit quantum gates}

\begin{figure}[t]
\centerline{\includegraphics[width=0.48\textwidth]{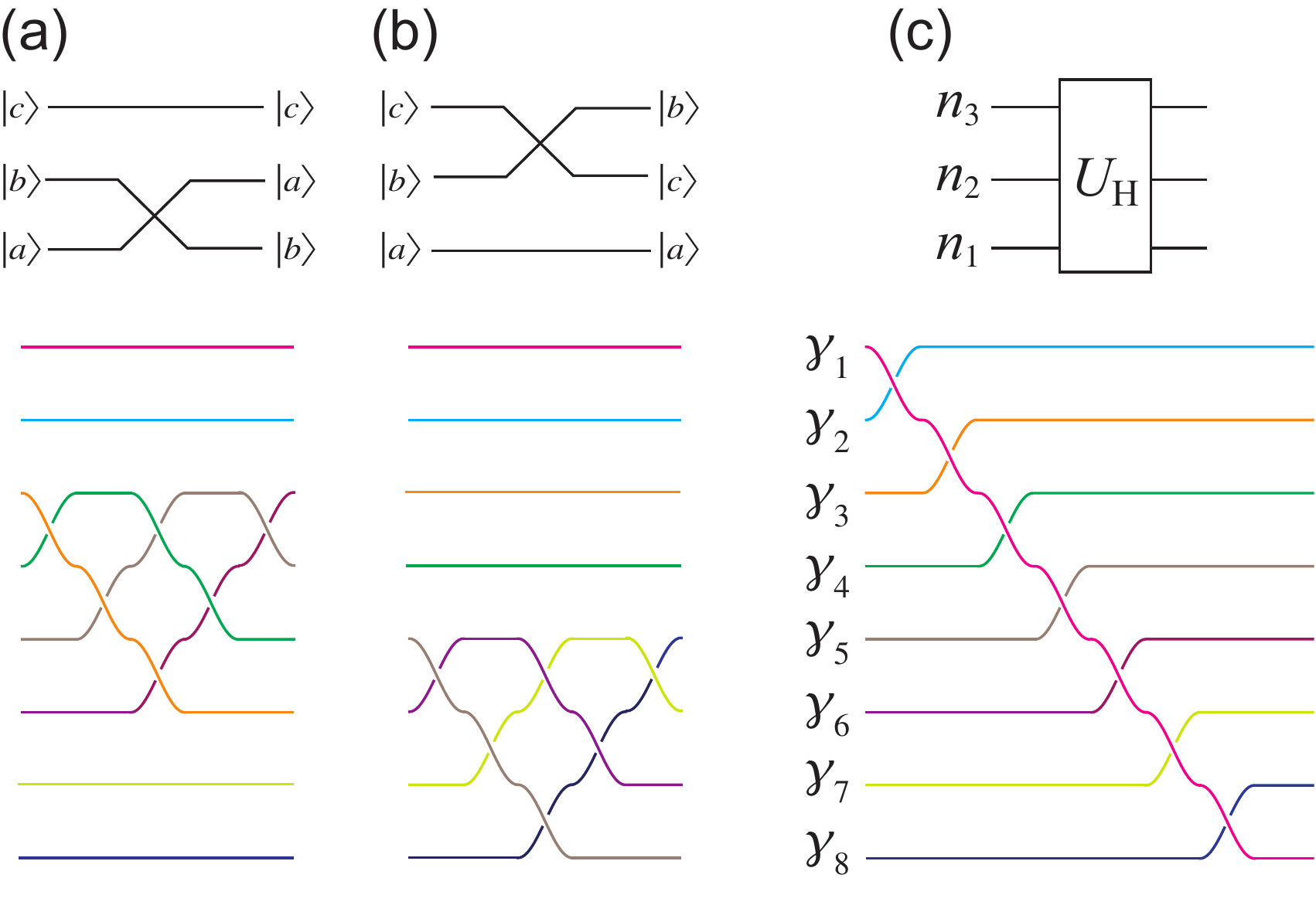}}
\caption{(a) and (b) $i$SWAP gate embedded to three qubit systems. (c)
Three-qubit Hadamard transformation. }
\label{FourBitH}
\end{figure}

We find that the three-qubit Hadamard transformation is generated as%
\begin{equation}
U_{\text{H}}^{\left( 3\right) }=-\mathcal{B}_{7}\mathcal{B}_{6}\mathcal{B}%
_{5}\mathcal{B}_{4}\mathcal{B}_{3}\mathcal{B}_{2}\mathcal{B}_{1}=\left( 
\begin{array}{cccccccc}
1 & 1 & 1 & 1 & 1 & 1 & 1 & 1 \\ 
1 & -1 & 1 & -1 & 1 & -1 & 1 & -1 \\ 
1 & -1 & -1 & 1 & 1 & -1 & -1 & 1 \\ 
1 & 1 & -1 & -1 & 1 & 1 & -1 & -1 \\ 
1 & -1 & -1 & 1 & -1 & 1 & 1 & -1 \\ 
1 & 1 & -1 & -1 & -1 & -1 & 1 & 1 \\ 
1 & 1 & 1 & 1 & -1 & -1 & -1 & -1 \\ 
1 & -1 & 1 & -1 & -1 & 1 & -1 & 1%
\end{array}%
\right) .
\end{equation}%
See Fig.\ref{FourBitH}(b). It is different from the cross-product of the
Hadamard gate%
\begin{equation}
U_{\text{H}}\otimes U_{\text{H}}\otimes U_{\text{H}}=\left( 
\begin{array}{cccccccc}
1 & 1 & 1 & 1 & 1 & 1 & 1 & 1 \\ 
1 & -1 & 1 & -1 & 1 & -1 & 1 & -1 \\ 
1 & 1 & -1 & -1 & 1 & 1 & -1 & -1 \\ 
1 & -1 & -1 & 1 & 1 & -1 & -1 & 1 \\ 
1 & 1 & 1 & 1 & -1 & -1 & -1 & -1 \\ 
1 & -1 & 1 & -1 & -1 & 1 & -1 & 1 \\ 
1 & 1 & -1 & -1 & -1 & -1 & 1 & 1 \\ 
1 & -1 & -1 & 1 & -1 & 1 & 1 & -1%
\end{array}%
\right) .
\end{equation}%
There is a relation%
\begin{equation}
U_{\text{H}}\otimes U_{\text{H}}\otimes U_{\text{H}}=-U_{3\text{P}}U_{\text{H%
}}^{\left( 3\right) },
\end{equation}%
where $U_{3\text{P}}$ is defined by%
\begin{equation}
U\equiv \left( 
\begin{array}{cccccccc}
1 & 0 & 0 & 0 & 0 & 0 & 0 & 0 \\ 
0 & 1 & 0 & 0 & 0 & 0 & 0 & 0 \\ 
0 & 0 & 0 & 1 & 0 & 0 & 0 & 0 \\ 
0 & 0 & 1 & 0 & 0 & 0 & 0 & 0 \\ 
0 & 0 & 0 & 0 & 0 & 0 & 1 & 0 \\ 
0 & 0 & 0 & 0 & 0 & 0 & 0 & 1 \\ 
0 & 0 & 0 & 0 & 0 & 1 & 0 & 0 \\ 
0 & 0 & 0 & 0 & 1 & 0 & 0 & 0%
\end{array}%
\right) .
\end{equation}%
It is impossible to generate the W state by braiding%
\begin{equation}
\left\vert \text{W}\right\rangle _{\text{logical}}=\frac{1}{\sqrt{3}}\left(
\left\vert 000\right\rangle _{\text{logical}}+\left\vert 010\right\rangle _{%
\text{logical}}+\left\vert 100\right\rangle _{\text{logical}}\right) ,
\end{equation}%
because the number of the nonzero terms of the W state is 3, which
contradicts the fact that the number of the nonzero terms must be $1$, $2$, $%
4$ and $8$ for three qubit states generated by braiding.

\subsection{$N$ logical qubits}

The braid representation of $2N+2$ Majorana fermions is equivalent to the $%
\pi /2$ rotation in SO$\left( 2N+2\right) $, suggested by the fact that
braid operators are represented by the Gamma matrices\cite%
{NayakWilczekS,GeorgievJSS}. The number of the braid group is given by\cite%
{ReadBraidS}

\begin{equation}
\left\vert \text{Image}\left( \mathcal{B}_{2n}\right) \right\vert =\left\{ 
\begin{array}{ccc}
2^{2n-1}\left( 2n\right) ! & \text{for} & n\text{=even} \\ 
2^{2n}\left( 2n\right) ! & \text{for} & n\text{=odd}%
\end{array}%
\right. .
\end{equation}

The $i$SWAP gate is embedded as%
\begin{equation}
I_{2}^{k-2}\otimes U_{i\text{SWAP}}\otimes I_{2}^{N-k}\propto \mathcal{B}%
_{2k+1}\mathcal{B}_{2k+2}\mathcal{B}_{2k+3}\mathcal{B}_{2k+1}\mathcal{B}%
_{2k+2}\mathcal{B}_{2k+1}.
\end{equation}

\subsubsection{Diagonal braiding}

We consider odd braiding defined by%
\begin{equation}
\mathcal{B}_{\text{odd}}\left( n_{1},n_{2},\cdots ,n_{k}\right) \equiv 
\mathcal{B}_{2n_{k}-1}\mathcal{B}_{2n_{k-1}-1}\cdots \mathcal{B}_{2n_{1}-1},
\end{equation}%
where $n_{k}$ is an integer satisfying $1\leq n_{k}\leq N+1$. They are
Abelian braiding because there are no adjacent braiding. Then, there are
only $4^{k}$ patterns. Especially, we consider odd double braiding defined by%
\begin{equation}
\left( \mathcal{B}_{\text{odd}}\right) ^{2}\equiv \left( i\mathcal{B}%
_{2n_{k}-1}^{2}\right) \left( i\mathcal{B}_{2n_{k-1}-1}^{2}\right) \cdots
\left( i\mathcal{B}_{2n_{1}-1}^{2}\right)
\end{equation}%
are interesting because they are identical to%
\begin{equation}
\left( \mathcal{B}_{\text{odd}}\right) ^{2}=\left( \sigma _{\text{Z}}\right)
^{m_{k}}\left( \sigma _{\text{Z}}\right) ^{m_{2}}\cdots \left( \sigma _{%
\text{Z}}\right) ^{m_{1}},
\end{equation}%
where $m_{k}=0,1$. Namely, every Pauli gates constructing from the Pauli Z
gate can be generated.

Next, we consider even braiding defined by%
\begin{equation}
\mathcal{B}_{\text{even}}\left( n_{1},n_{2},\cdots ,n_{k}\right) \equiv 
\mathcal{B}_{2n_{k}}\mathcal{B}_{2n_{k-1}}\cdots \mathcal{B}_{2n_{1}}.
\end{equation}%
They are also the Abelian braiding, where each braiding commutes each other.
We also consider even double braiding defined by%
\begin{equation}
\left( \mathcal{B}_{\text{even}}\right) ^{2}\equiv \left( i\mathcal{B}%
_{2n_{k}}^{2}\right) \left( i\mathcal{B}_{2n_{k-1}}^{2}\right) \cdots \left(
i\mathcal{B}_{2n_{1}}^{2}\right) .
\end{equation}%
On the other hand, it is impossible to construct the Pauli X gate except for
the first qubit. See Eq.(\ref{BPauliX}) in the main text.

\subsubsection{Hadamard transformation}

The Hadamard transformation is used for the initial process of various
quantum algorithm such as the Kitaev phase estimation algorithm, the Deutsch algorithm, the Deutsch-Jozsa
algorithm, the Simon algorithm, the
Bernstein-Vazirani algorithm, the Grover algorithm
and the Shor algorithm. It is generated by the braiding%
\begin{equation}
U_{\text{H}}^{\left( N\right) }\propto \mathcal{B}_{2N+1}\mathcal{B}%
_{2N}\cdots \mathcal{B}_{2}\mathcal{B}_{1}.
\end{equation}%
The equal-coefficient state is generated as%
\begin{equation}
U_{\text{H}}^{\left( N\right) }\left\vert 0,0\right\rangle _{\text{logical}%
}\propto \sum_{j=1}^{2^{N}}\left\vert j\right\rangle _{\text{logical}},
\end{equation}%
where $\left\vert j\right\rangle _{\text{logical}}$ is the decimal
representation of the qubit.

\section{$2N$-body unitary evolution}

\subsection{Quantum gates for one logical qubit}

The $2$-body Majorana operator $\mathcal{B}_{1}\left( \theta \right) $ is
written in terms of fermion operators,%
\begin{equation}
\mathcal{B}_{1}\left( \theta \right) =\cos \theta +\gamma _{2}\gamma
_{1}\sin \theta =\left( \cos \theta +\left( ic_{1}^{\dagger
}c_{1}-ic_{1}c_{1}^{\dagger }\right) \sin \theta \right) ,
\end{equation}%
which operates on two physical qubits (\ref{PsiTwo}) as%
\begin{equation}
\mathcal{B}_{1}\left( \theta \right) \Psi _{\text{physical}}=\left( 
\begin{array}{cccc}
e^{-i\theta } & 0 & 0 & 0 \\ 
0 & e^{i\theta } & 0 & 0 \\ 
0 & 0 & e^{-i\theta } & 0 \\ 
0 & 0 & 0 & e^{i\theta }%
\end{array}%
\right) \left( 
\begin{array}{c}
\left\vert 0,0\right\rangle \\ 
\left\vert 0,1\right\rangle \\ 
\left\vert 1,0\right\rangle \\ 
\left\vert 1,1\right\rangle%
\end{array}%
\right) _{\text{physical}}.
\end{equation}%
Taking the even parity basis, the action is%
\begin{equation}
\mathcal{B}_{1}\left( \theta \right) \Psi _{\text{logical}}=e^{-i\theta
}\left( 
\begin{array}{cc}
1 & 0 \\ 
0 & e^{2i\theta }%
\end{array}%
\right) \left( 
\begin{array}{c}
\left\vert 0\right\rangle \\ 
\left\vert 1\right\rangle%
\end{array}%
\right) _{\text{logical}}.  \label{B1US2}
\end{equation}%
It is the arbitrary phase-shift gate. Especially, by setting $\theta =\pi /8$%
, the T gate is constructed%
\begin{equation}
U_{\text{T}}\equiv \text{diag.}\left( 1,e^{i\pi /4}\right) .
\end{equation}%
It is identical to the rotation along the $z$ axis%
\begin{equation}
\mathcal{B}_{1}\left( \theta \right) =R_{z}\left( 2\theta \right) ,
\label{Uzt}
\end{equation}%
with%
\begin{equation}
R_{z}\left( \theta \right) \equiv \exp \left[ -i\frac{\theta }{2}\sigma _{z}%
\right] =\left( 
\begin{array}{cc}
e^{-i\theta /2} & 0 \\ 
0 & e^{i\theta /2}%
\end{array}%
\right) .
\end{equation}

The operator $\mathcal{B}_{2}$ is written in terms of fermion operators, 
\begin{equation}
\mathcal{B}_{2}\left( \theta \right) =\cos \theta +\gamma _{3}\gamma
_{2}\sin \theta =\cos \theta +\left( ic_{2}c_{1}^{\dagger }+ic_{2}^{\dagger
}c_{1}^{\dagger }-ic_{2}c_{1}-ic_{2}^{\dagger }c_{1}\right) \sin \theta ,
\end{equation}%
which operates on two physical qubits (\ref{PsiTwo}) as\cite{IvanovS},%
\begin{equation}
\mathcal{B}_{2}\left( \theta \right) \Psi _{\text{physical}}=\left( 
\begin{array}{cccc}
\cos \theta & 0 & 0 & -i\sin \theta \\ 
0 & \cos \theta & -i\sin \theta & 0 \\ 
0 & -i\sin \theta & \cos \theta & 0 \\ 
-i\sin \theta & 0 & 0 & \cos \theta%
\end{array}%
\right) \left( 
\begin{array}{c}
\left\vert 0,0\right\rangle \\ 
\left\vert 0,1\right\rangle \\ 
\left\vert 1,0\right\rangle \\ 
\left\vert 1,1\right\rangle%
\end{array}%
\right) _{\text{physical}}.
\end{equation}%
In the even parity basis, the action is%
\begin{equation}
\mathcal{B}_{2}\left( \theta \right) =\left( 
\begin{array}{cc}
\cos \theta & -i\sin \theta \\ 
-i\sin \theta & \cos \theta%
\end{array}%
\right) \equiv R_{x}\left( 2\theta \right) ,
\end{equation}%
which is identical to the rotation along the $x$ axis%
\begin{equation}
\mathcal{B}_{2}\left( \theta \right) =R_{x}\left( 2\theta \right) ,
\label{Uxt}
\end{equation}%
with%
\begin{equation}
R_{x}\left( \theta \right) \equiv \exp \left[ -i\frac{\theta }{2}\sigma _{x}%
\right] =\left( 
\begin{array}{cc}
\cos \frac{\theta }{2} & -i\sin \frac{\theta }{2} \\ 
-i\sin \frac{\theta }{2} & \cos \frac{\theta }{2}%
\end{array}%
\right) .
\end{equation}

The operator $\mathcal{B}_{3}\left( \theta \right) $ is written in terms of
fermion operators%
\begin{equation}
\mathcal{B}_{3}\left( \theta \right) =\cos \theta +\gamma _{4}\gamma
_{3}\sin \theta =\cos \theta +\left( ic_{2}^{\dagger
}c_{2}-ic_{2}c_{2}^{\dagger }\right) \sin \theta ,
\end{equation}%
which operates on two physical qubits (\ref{PsiTwo}) as\cite{IvanovS}%
\begin{equation}
\mathcal{B}_{3}\left( \theta \right) \Psi _{\text{physical}}=\left( 
\begin{array}{cccc}
e^{-i\theta } & 0 & 0 & 0 \\ 
0 & e^{-i\theta } & 0 & 0 \\ 
0 & 0 & e^{i\theta } & 0 \\ 
0 & 0 & 0 & e^{i\theta }%
\end{array}%
\right) \left( 
\begin{array}{c}
\left\vert 0,0\right\rangle \\ 
\left\vert 0,1\right\rangle \\ 
\left\vert 1,0\right\rangle \\ 
\left\vert 1,1\right\rangle%
\end{array}%
\right) _{\text{physical}}.
\end{equation}%
In the even parity basis, the action is the same as (\ref{B1US2}),%
\begin{equation}
\mathcal{B}_{3}\left( \theta \right) =e^{-i\theta }\left( 
\begin{array}{cc}
1 & 0 \\ 
0 & e^{2i\theta }%
\end{array}%
\right) \left( 
\begin{array}{c}
\left\vert 0\right\rangle \\ 
\left\vert 1\right\rangle%
\end{array}%
\right) _{\text{logical}}.
\end{equation}

The rotation along the $y$ axis defined by

\begin{equation}
R_{y}\left( \theta \right) \equiv \exp \left[ -i\frac{\theta }{2}\sigma _{y}%
\right]
\end{equation}%
is realized by the sequential operations%
\begin{equation}
R_{y}\left( \theta \right) =R_{z}\left( \frac{\pi }{2}\right) R_{x}\left(
\theta \right) R_{z}\left( -\frac{\pi }{2}\right).
\end{equation}

\subsection{Three physical qubits}

Next, we study the six Majorana fermion system. The explicit actions on the
physical qubits are%
\begin{align}
\mathcal{B}_{1}\left( \theta \right) & =I_{2}\otimes I_{2}\otimes
R_{z}\left( 2\theta \right) ,  \notag \\
\mathcal{B}_{2}\left( \theta \right) & =I_{2}\otimes U_{xx}\left( \theta
\right) ,  \notag \\
\mathcal{B}_{3}\left( \theta \right) & =I_{2}\otimes R_{z}\left( 2\theta
\right) \otimes I_{2},  \notag \\
\mathcal{B}_{4}\left( \theta \right) & =U_{xx}\otimes I_{2},  \notag \\
\mathcal{B}_{5}\left( \theta \right) & =R_{z}\left( 2\theta \right) \otimes
I_{2}\otimes I_{2}.
\end{align}

\subsection{Two logical qubits}

Two logical qubits are constructed from three physical qubits by taking the
even parity basis. The action of $\mathcal{B}_{1}\left( \theta \right) $ to
the logical qubit is%
\begin{equation}
\mathcal{B}_{1}\left( \theta \right) =\text{diag.}\left( e^{-i\theta
},e^{i\theta },e^{i\theta },e^{-i\theta }\right) ,
\end{equation}%
which is identical to the ZZ interaction%
\begin{equation}
\mathcal{B}_{1}\left( \theta \right) =U_{zz}\left( 2\theta \right) ,
\end{equation}%
with%
\begin{equation}
U_{zz}\left( \theta \right) \equiv \exp \left[ -i\frac{\theta }{2}\sigma
_{z}\otimes \sigma _{z}\right] .
\end{equation}

The action of $\mathcal{B}_{4}\left( \theta \right) $ to the logical qubit is%
\begin{equation}
\mathcal{B}_{4}\left( \theta \right) =\left( 
\begin{array}{cccc}
\cos \theta & 0 & 0 & -i\sin \theta \\ 
0 & \cos \theta & -i\sin \theta & 0 \\ 
0 & -i\sin \theta & \cos \theta & 0 \\ 
-i\sin \theta & 0 & 0 & \cos \theta%
\end{array}%
\right) ,
\end{equation}%
which is identical to the xx interaction%
\begin{equation}
\mathcal{B}_{4}\left( \theta \right) =U_{xx}\left( 2\theta \right) ,
\end{equation}%
with%
\begin{equation}
U_{xx}\left( \theta \right) \equiv \exp \left[ -i\frac{\theta }{2}\sigma
_{x}\otimes \sigma _{x}\right] .
\end{equation}

The action of $\mathcal{B}_{3}\left( \theta \right) $ and $\mathcal{B}%
_{5}\left( \theta \right) $ to the logical qubit is%
\begin{eqnarray}
\mathcal{B}_{3}\left( \theta \right) &=&\text{diag.}\left( e^{-i\theta
},e^{i\theta },e^{-i\theta },e^{i\theta }\right) =I_{2}\otimes U_{z}\left(
\theta \right) ,  \notag \\
\mathcal{B}_{5}\left( \theta \right) &=&\text{diag.}\left( e^{-i\theta
},e^{-i\theta },e^{i\theta },e^{i\theta }\right) =U_{z}\left( \theta \right)
\otimes I_{2}.
\end{eqnarray}

The action of $\mathcal{B}_{2}\left( \theta \right) $ to the logical qubit is%
\begin{equation}
\mathcal{B}_{2}\left( \theta \right) =\left( 
\begin{array}{cccc}
\cos \theta & -i\sin \theta & 0 & 0 \\ 
-i\sin \theta & \cos \theta & 0 & 0 \\ 
0 & 0 & \cos \theta & -i\sin \theta \\ 
0 & 0 & -i\sin \theta & \cos \theta%
\end{array}%
\right) ,
\end{equation}%
which is rewritten in the form of%
\begin{eqnarray}
\mathcal{B}_{2}\left( \theta \right) &=&I_{2}\otimes R_{x}\left( 2\theta
\right) ,  \notag \\
\mathcal{B}_{2345}\left( \theta \right) &=&R_{x}\left( 2\theta \right)
\otimes I_{2}.
\end{eqnarray}

\subsection{Controlled phase-shift gate}

We find%
\begin{equation}
\mathcal{B}_{6}\left( \theta _{3}\right) \mathcal{B}_{3}\left( \theta
_{2}\right) \mathcal{B}_{1}\left( \theta _{1}\right) =\text{diag.}\left(
e^{-i\left( \theta _{1}+\theta _{2}+\theta _{3}\right) },e^{i\left( \theta
_{1}+\theta _{2}-\theta _{3}\right) },e^{i\left( \theta _{1}-\theta
_{2}+\theta _{3}\right) },e^{i\left( -\theta _{1}+\theta _{2}+\theta
_{3}\right) }\right) .
\end{equation}%
The controlled phase-shift gate with arbitrary phase is constructed by
setting $\theta _{1}=-\theta _{2}=-\theta _{3}=-\theta $,%
\begin{equation}
\mathcal{B}_{5}\left( -\theta \right) \mathcal{B}_{3}\left( \theta \right) 
\mathcal{B}_{1}\left( \theta \right) =\text{diag.}\left( e^{-i\theta
},e^{-i\theta },e^{-i\theta },e^{3i\theta }\right) =e^{-i\theta }\text{diag.}%
\left( 1,1,1,e^{4i\theta }\right) .
\end{equation}%
Especially, the CZ gate is constructed by setting $\theta =\pi /4.$

\subsection{\textit{Controlled-unitary gate}}

It is known that the controlled unitary gate is constructed as\cite{BarencoS}%
\begin{equation}
U_{\text{C-}U}=\left( I_{2}\otimes U_{A}\right) U_{\text{CNOT}}\left(
I_{2}\otimes U_{B}\right) U_{\text{CNOT}}\left( I_{2}\otimes U_{C}\right)
\end{equation}%
with%
\begin{equation}
U_{A}\equiv R_{z}\left( \beta \right) R_{y}\left( \frac{\gamma }{2}\right)
,\qquad U_{B}\equiv R_{y}\left( -\frac{\gamma }{2}\right) R_{z}\left( -\frac{%
\beta +\delta }{2}\right) ,\qquad U_{C}\equiv R_{z}\left( \frac{\delta
-\beta }{2}\right) ,
\end{equation}%
because%
\begin{equation}
U_{A}U_{B}U_{C}=I_{4}
\end{equation}%
and%
\begin{equation}
U_{A}XU_{B}XU_{C}=R_{z}\left( \beta \right) R_{y}\left( \gamma \right)
R_{z}\left( \delta \right) =U_{\text{1bit}}.
\end{equation}%
In the Majorana system, the basic rotations are not along the $y$ axis but
the $x$ axis. The similar decomposition is possible only by using the
rotations along the $z$ and $x$ axes as%
\begin{equation}
U_{A}=R_{z}\left( \beta +\frac{\pi }{2}\right) R_{x}\left( \frac{\gamma }{2}%
\right) ,\qquad U_{B}=R_{x}\left( -\frac{\gamma }{2}\right) R_{z}\left( -%
\frac{\beta +\delta }{2}\right) ,\qquad U_{C}=R_{z}\left( \frac{\delta
-\beta -\pi }{2}\right) .
\end{equation}%
The proof is similar. First, we have%
\begin{equation*}
U_{A}U_{B}U_{C}=1,
\end{equation*}%
where we have used the relation%
\begin{equation}
R_{j}\left( \theta _{1}\right) R_{j}\left( \theta _{2}\right) =R_{j}\left(
\theta _{1}+\theta _{2}\right)
\end{equation}%
for $j=x,y$ and $z$. Next, we have%
\begin{equation*}
U_{A}XU_{B}XU_{C}=U_{\text{1bit}},
\end{equation*}%
where we have used the relation%
\begin{equation}
R\left( \theta \right) X=XR\left( -\theta \right) .
\end{equation}%
Hence, the controlled unitary gate is implemented by $2$-body Majorana
interaction.

\subsection{Four physical qubits}

We consider eight Majorana fermion system The explicit actions on four
physical qubits are given by%
\begin{align}
\mathcal{B}_{1}& =I_{8}\otimes R_{z}\left( 2\theta \right) ,  \notag \\
\mathcal{B}_{2}& =I_{4}\otimes U_{xx}\left( \theta \right) ,  \notag \\
\mathcal{B}_{3}& =I_{4}\otimes R_{z}\left( 2\theta \right) \otimes I_{2}, 
\notag \\
\mathcal{B}_{4}& =I_{2}\otimes U_{xx}\left( \theta \right) \otimes I_{2}, 
\notag \\
\mathcal{B}_{5}& =I_{2}\otimes R_{z}\left( 2\theta \right) \otimes I_{4}, 
\notag \\
\mathcal{B}_{6}& =U_{xx}\left( \theta \right) \otimes I_{4},  \notag \\
\mathcal{B}_{7}& =R_{z}\left( 2\theta \right) \otimes I_{8}.
\end{align}%
We summarize results on constructing full set of Pauli Z gate for three
logical qubits in \ the following table: 
\begin{equation}
\begin{tabular}{|l|l|l|}
\hline
& 4 physical qubits & 3 logical qubits \\ \hline
$\quad \mathcal{B}_{12}\left( \theta \right) $ & $\quad \exp \left[ -i\theta
I_{8}\otimes \sigma _{z}\right] $ & $\quad \exp \left[ -i\theta \sigma
_{z}\otimes \sigma _{z}\otimes \sigma _{z}\right] \quad $ \\ \hline
$\quad \mathcal{B}_{34}\left( \theta \right) $ & $\quad \exp \left[ -i\theta
I_{4}\otimes \sigma _{z}\otimes I_{2}\right] $ & $\quad \exp \left[ -i\theta
I_{4}\otimes \sigma _{z}\right] $ \\ \hline
$\quad \mathcal{B}_{56}\left( \theta \right) $ & $\quad \exp \left[ -i\theta
I_{2}\otimes \sigma _{z}\otimes I_{4}\right] $ & $\quad \exp \left[ -i\theta
I_{2}\otimes \sigma _{z}\otimes I_{2}\right] $ \\ \hline
$\quad \mathcal{B}_{78}\left( \theta \right) $ & $\quad \exp \left[ -i\theta
\sigma _{z}\otimes I_{8}\right] $ & $\quad \exp \left[ -i\theta \sigma
_{z}\otimes I_{4}\right] $ \\ \hline
$\quad \mathcal{B}_{1234}^{\left( 4\right) }\left( \theta \right) $ & $\quad
\exp \left[ -i\theta I_{4}\otimes \sigma _{z}\otimes \sigma _{z}\right] $ & $%
\quad \exp \left[ -i\theta \sigma _{z}\otimes \sigma _{z}\otimes I_{2}\right]
$ \\ \hline
$\quad \mathcal{B}_{1256}^{\left( 4\right) }\left( \theta \right) $ & $\quad
\exp \left[ -i\theta I_{2}\otimes \sigma _{z}\otimes I_{2}\otimes \sigma _{z}%
\right] $ & $\quad \exp \left[ -i\theta \sigma _{z}\otimes I_{2}\otimes
\sigma _{z}\right] $ \\ \hline
$\quad \mathcal{B}_{1278}^{\left( 4\right) }\left( \theta \right) $ & $\quad
\exp \left[ -i\theta \sigma _{z}\otimes I_{2}\otimes I_{2}\otimes \sigma _{z}%
\right] $ & $\quad \exp \left[ -i\theta I_{2}\otimes \sigma _{z}\otimes
\sigma _{z}\right] $ \\ \hline
$\quad \mathcal{B}_{3456}^{\left( 4\right) }\left( \theta \right) $ & $\quad
\exp \left[ -i\theta I_{2}\otimes \sigma _{z}\otimes \sigma _{z}\otimes I_{2}%
\right] $ & $\quad \exp \left[ -i\theta I_{2}\otimes \sigma _{z}\otimes
\sigma _{z}\right] $ \\ \hline
$\quad \mathcal{B}_{3478}^{\left( 4\right) }\left( \theta \right) $ & $\quad
\exp \left[ -i\theta \sigma _{z}\otimes I_{2}\otimes \sigma _{z}\otimes I_{2}%
\right] $ & $\quad \exp \left[ -i\theta \sigma _{z}\otimes I_{2}\otimes
\sigma _{z}\right] $ \\ \hline
$\quad \mathcal{B}_{5678}^{\left( 4\right) }\left( \theta \right) $ & $\quad
\exp \left[ -i\theta \sigma _{z}\otimes \sigma _{z}\otimes I_{2}\otimes I_{2}%
\right] $ & $\quad \exp \left[ -i\theta \sigma _{z}\otimes \sigma
_{z}\otimes I_{2}\right] $ \\ \hline
$\quad \mathcal{B}_{123456}^{\left( 6\right) }\left( \theta \right) $ & $%
\quad \exp \left[ -i\theta I_{2}\otimes \sigma _{z}\otimes \sigma
_{z}\otimes \sigma _{z}\right] $ & $\quad \exp \left[ -i\theta \sigma
_{z}\otimes I_{4}\right] $ \\ \hline
$\quad \mathcal{B}_{123478}^{\left( 6\right) }\left( \theta \right) $ & $%
\quad \exp \left[ -i\theta \sigma _{z}\otimes I_{2}\otimes \sigma
_{z}\otimes \sigma _{z}\right] $ & $\quad \exp \left[ -i\theta I_{2}\otimes
\sigma _{z}\otimes I_{2}\right] $ \\ \hline
$\quad \mathcal{B}_{125678}^{\left( 6\right) }\left( \theta \right) $ & $%
\quad \exp \left[ -i\theta \sigma _{z}\otimes \sigma _{z}\otimes
I_{2}\otimes \sigma _{z}\right] $ & $\quad \exp \left[ -i\theta I_{4}\otimes
\sigma _{z}\right] $ \\ \hline
$\quad \mathcal{B}_{345678}^{\left( 6\right) }\left( \theta \right) $ & $%
\quad \exp \left[ -i\theta \sigma _{z}\otimes \sigma _{z}\otimes \sigma
_{z}\otimes I_{2}\right] $ & $\quad \exp \left[ -i\theta \sigma _{z}\otimes
\sigma _{z}\otimes \sigma _{z}\right] $ \\ \hline
$\quad \mathcal{B}_{12345678}^{\left( 8\right) }\left( \theta \right) \quad $
& $\quad \exp \left[ -i\theta \sigma _{z}\otimes \sigma _{z}\otimes \sigma
_{z}\otimes \sigma _{z}\right] \quad $ & $\quad \exp \left[ -i\theta I_{8}%
\right] $ \\ \hline
\end{tabular}
\label{table}
\end{equation}

We have the identical logical qubits: 
\begin{eqnarray}
\mathcal{B}_{12}\left( \theta \right) &\simeq &\mathcal{B}_{345678}\left(
\theta \right) ,\qquad \mathcal{B}_{34}\left( \theta \right) \simeq \mathcal{%
B}_{125678}\left( \theta \right) ,\qquad \mathcal{B}_{56}\left( \theta
\right) \simeq \mathcal{B}_{123478}\left( \theta \right) ,\qquad \mathcal{B}%
_{78}\left( \theta \right) \simeq \mathcal{B}_{123456}\left( \theta \right) ,
\notag \\
\mathcal{B}_{1234}\left( \theta \right) &\simeq &\mathcal{B}_{5678}\left(
\theta \right) ,\qquad \mathcal{B}_{1256}\left( \theta \right) \simeq 
\mathcal{B}_{3478}\left( \theta \right) ,\qquad \mathcal{B}_{1278}\left(
\theta \right) \simeq \mathcal{B}_{3456}\left( \theta \right) .
\label{CompB}
\end{eqnarray}

\subsection{Three logical qubits}

Three logical qubits are constructed from four physical qubits by taking the
even parity basis. 
\begin{eqnarray}
\mathcal{B}_{12} &=&\text{diag.}\left( e^{-i\theta },e^{i\theta },e^{i\theta
},e^{-i\theta },e^{i\theta },e^{-i\theta },e^{-i\theta },e^{i\theta }\right)
=\exp \left[ -i\theta \sigma _{z}\otimes \sigma _{z}\otimes \sigma _{z}%
\right] ,  \notag \\
\mathcal{B}_{23} &=&I_{2}\otimes I_{2}\otimes R_{x}\left( 2\theta \right) , 
\notag \\
\mathcal{B}_{34} &=&I_{2}\otimes I_{2}\otimes R_{z}\left( 2\theta \right)
=\exp \left[ -i\theta I_{2}\otimes I_{2}\otimes \sigma _{z}\right] ,  \notag
\\
\mathcal{B}_{45} &=&I_{2}\otimes U_{xx}\left( 2\theta \right) ,  \notag \\
\mathcal{B}_{56} &=&I_{2}\otimes R_{z}\left( 2\theta \right) \otimes
I_{2}=\exp \left[ -i\theta I_{2}\otimes \sigma _{z}\otimes I_{2}\right] , 
\notag \\
\mathcal{B}_{67} &=&U_{xx}\left( 2\theta \right) \otimes I_{2},  \notag \\
\mathcal{B}_{78} &=&R_{z}\left( 2\theta \right) \otimes I_{4}.
\end{eqnarray}%
We find that controlled-controlled phase shift gate cannot be implemented
only by diagonal braiding. It is proved by counting the number of the
degrees of freedom. We need to tune 7 parameters for the diagonal quantum
gates. On the other hand, there are only three independent angle because the
diagonal operators are $\mathcal{B}_{1}$, $\mathcal{B}_{3}$ and $\mathcal{B}%
_{5}$. Hence, it is impossible to construct controlled-controlled phase
shift gate in general. However, this problem is solved by introducing
many-body Majorana interaction, 
\begin{equation}
U_{\text{CC}\phi }=\mathcal{B}_{12}\left( \frac{\phi }{8}\right) \mathcal{B}%
_{34}\left( \frac{\phi }{8}\right) \mathcal{B}_{56}\left( \frac{\phi }{8}%
\right) \mathcal{B}_{78}\left( \frac{\phi }{8}\right) \mathcal{B}%
_{1234}^{\left( 4\right) }\left( -\frac{\phi }{8}\right) \mathcal{B}%
_{1278}^{\left( 4\right) }\left( -\frac{\phi }{8}\right) \mathcal{B}%
_{1256}^{\left( 4\right) }\left( -\frac{\phi }{8}\right) .
\end{equation}%
Especially, the CCZ gate is constructed as follows%
\begin{equation}
U_{\text{CCZ}}=\mathcal{B}_{12}\left( \frac{\pi }{8}\right) \mathcal{B}%
_{34}\left( \frac{\pi }{8}\right) \mathcal{B}_{56}\left( \frac{\pi }{8}%
\right) \mathcal{B}_{78}\left( \frac{\pi }{8}\right) \mathcal{B}%
_{1234}^{\left( 4\right) }\left( -\frac{\pi }{8}\right) \mathcal{B}%
_{1278}^{\left( 4\right) }\left( -\frac{\pi }{8}\right) \mathcal{B}%
_{1256}^{\left( 4\right) }\left( -\frac{\pi }{8}\right) .
\end{equation}%
The Toffoli gate is constructed by applying the Hadamard gate to the CCZ
gate as in%
\begin{equation}
U_{\text{Toffloi}}=\left( I_{4}\otimes U_{\text{H}}\right) U_{\text{CCZ}%
}\left( I_{4}\otimes U_{\text{H}}\right) .
\end{equation}%
See Fig.\ref{Fredkin}(a).

The Fredkin gate is constructed by sequential applications of three Toffoli
gates as in%
\begin{equation}
U_{\text{Fredkin}}=U_{\text{Toffoli}}^{\left( 3,2\right) \rightarrow 1}U_{%
\text{Toffoli}}^{\left( 3,1\right) \rightarrow 2}U_{\text{Toffoli}}^{\left(
3,2\right) \rightarrow 1}.
\end{equation}%
See Fig.\ref{Fredkin}(b).

The CZ gate in three qubits are embedded as%
\begin{eqnarray}
U_{\text{CZ}}^{3\rightarrow 2} &=&\mathcal{B}_{56}\left( \frac{\pi }{4}%
\right) \mathcal{B}_{78}\left( \frac{\pi }{4}\right) \mathcal{B}%
_{1234}^{\left( 4\right) }\left( -\frac{\pi }{4}\right) ,  \notag \\
U_{\text{CZ}}^{3\rightarrow 1} &=&\mathcal{B}_{34}\left( \frac{\pi }{4}%
\right) \mathcal{B}_{78}\left( \frac{\pi }{4}\right) \mathcal{B}%
_{1234}^{\left( 4\right) }\left( -\frac{\pi }{4}\right) ,  \notag \\
U_{\text{CZ}}^{2\rightarrow 1} &=&\mathcal{B}_{34}\left( \frac{\pi }{4}%
\right) \mathcal{B}_{56}\left( \frac{\pi }{4}\right) \mathcal{B}%
_{1278}^{\left( 4\right) }\left( -\frac{\pi }{4}\right) .
\end{eqnarray}

\subsection{Logical 4 qubits}

We summarize results on constructing full set of Pauli Z gate for four
logical qubits in \ the following table:%
\begin{equation}
\begin{tabular}{|l|l|}
\hline
& 4 logical qubits \\ \hline
$\quad \mathcal{B}_{12}\left( \theta \right) $ & $\quad \exp \left[ -i\theta
\sigma _{z}\otimes \sigma _{z}\otimes \sigma _{z}\otimes \sigma _{z}\right]
\quad $ \\ \hline
$\quad \mathcal{B}_{34}\left( \theta \right) $ & $\quad \exp \left[ -i\theta
I_{8}\otimes \sigma _{z}\right] $ \\ \hline
$\quad \mathcal{B}_{56}\left( \theta \right) $ & $\quad \exp \left[ -i\theta
I_{4}\otimes \sigma _{z}\otimes I_{2}\right] $ \\ \hline
$\quad \mathcal{B}_{78}\left( \theta \right) $ & $\quad \exp \left[ -i\theta
I_{2}\otimes \sigma _{z}\otimes I_{4}\right] $ \\ \hline
$\quad \mathcal{B}_{90}\left( \theta \right) $ & $\quad \exp \left[ -i\theta
\sigma _{z}\otimes I_{8}\right] $ \\ \hline
$\quad \mathcal{B}_{1234}^{\left( 4\right) }\left( \theta \right) $ & $\quad
\exp \left[ -i\theta \sigma _{z}\otimes \sigma _{z}\otimes \sigma
_{z}\otimes I_{2}\right] $ \\ \hline
$\quad \mathcal{B}_{1256}^{\left( 4\right) }\left( \theta \right) $ & $\quad
\exp \left[ -i\theta \sigma _{z}\otimes \sigma _{z}\otimes I_{2}\otimes
\sigma _{z}\right] $ \\ \hline
$\quad \mathcal{B}_{1278}^{\left( 4\right) }\left( \theta \right) $ & $\quad
\exp \left[ -i\theta \sigma _{z}\otimes I_{2}\otimes \sigma _{z}\otimes
\sigma _{z}\right] $ \\ \hline
$\quad \mathcal{B}_{1290}^{\left( 4\right) }\left( \theta \right) $ & $\quad
\exp \left[ -i\theta I_{2}\otimes \sigma _{z}\otimes \sigma _{z}\otimes
\sigma _{z}\right] $ \\ \hline
$\quad \mathcal{B}_{3456}^{\left( 4\right) }\left( \theta \right) $ & $\quad
\exp \left[ -i\theta I_{2}\otimes I_{2}\otimes \sigma _{z}\otimes \sigma _{z}%
\right] $ \\ \hline
$\quad \mathcal{B}_{3478}^{\left( 4\right) }\left( \theta \right) $ & $\quad
\exp \left[ -i\theta I_{2}\otimes \sigma _{z}\otimes I_{2}\otimes \sigma _{z}%
\right] $ \\ \hline
$\quad \mathcal{B}_{3490}^{\left( 4\right) }\left( \theta \right) $ & $\quad
\exp \left[ -i\theta \sigma _{z}\otimes I_{4}\otimes \sigma _{z}\right] $ \\ 
\hline
$\quad \mathcal{B}_{5678}^{\left( 4\right) }\left( \theta \right) $ & $\quad
\exp \left[ -i\theta I_{2}\otimes \sigma _{z}\otimes \sigma _{z}\otimes I_{2}%
\right] $ \\ \hline
$\quad \mathcal{B}_{5690}^{\left( 4\right) }\left( \theta \right) $ & $\quad
\exp \left[ -i\theta \sigma _{z}\otimes I_{2}\otimes \sigma _{z}\otimes I_{2}%
\right] $ \\ \hline
$\quad \mathcal{B}_{7890}^{\left( 4\right) }\left( \theta \right) \quad $ & $%
\quad \exp \left[ -i\theta \sigma _{z}\otimes \sigma _{z}\otimes I_{4}\right]
$ \\ \hline
\end{tabular}%
\end{equation}%
where $0$ is an abbreviation of $10$. The CCC$\phi $ gate is explicitly
constructed as%
\begin{eqnarray}
&&U_{\text{C}^{3}\phi }=\mathcal{B}_{34}\left( \frac{\phi }{16}\right) 
\mathcal{B}_{56}\left( \frac{\phi }{16}\right) \mathcal{B}_{78}\left( \frac{%
\phi }{16}\right) \mathcal{B}_{90}\left( \frac{\phi }{16}\right) \mathcal{B}%
_{1234}^{\left( 4\right) }\left( \frac{\phi }{16}\right) \mathcal{B}%
_{1256}^{\left( 4\right) }\left( \frac{\phi }{16}\right) \mathcal{B}%
_{1278}^{\left( 4\right) }\left( \frac{\phi }{16}\right) \mathcal{B}%
_{1290}^{\left( 4\right) }\left( \frac{\phi }{16}\right)   \notag \\
&&\mathcal{B}_{3456}^{\left( 4\right) }\left( -\frac{\phi }{16}\right) 
\mathcal{B}_{3478}^{\left( 4\right) }\left( -\frac{\phi }{16}\right) 
\mathcal{B}_{3490}^{\left( 4\right) }\left( -\frac{\phi }{16}\right) 
\mathcal{B}_{5678}^{\left( 4\right) }\left( \frac{\phi }{16}\right) \mathcal{%
B}_{5690}^{\left( 4\right) }\left( -\frac{\phi }{16}\right) \mathcal{B}%
_{7890}^{\left( 4\right) }\left( -\frac{\phi }{16}\right) \mathcal{B}%
_{12}^{\left( 4\right) }\left( -\frac{\phi }{16}\right) .
\end{eqnarray}

\subsection{Generalized braid group relation}

We consider the case $\theta =\pi /4$. The Artin braid group relation reads%
\cite{ArtinS},%
\begin{equation}
\mathcal{B}_{\alpha }\mathcal{B}_{\beta }=\mathcal{B}_{\beta }\mathcal{B}%
_{\alpha }\qquad \text{for\quad }\left\vert \alpha -\beta \right\vert \geq
2,\qquad \mathcal{B}_{\alpha }\mathcal{B}_{\alpha +1}\mathcal{B}_{\alpha }=%
\mathcal{B}_{\alpha +1}\mathcal{B}_{\alpha }\mathcal{B}_{\alpha +1}.
\end{equation}%
It is identical to the extraspecial 2 group\cite{FrankoS}%
\begin{eqnarray}
M_{\alpha }^{2} &=&-1,\qquad M_{\alpha }M_{\alpha +1}=-M_{\alpha
+1}M_{\alpha },  \notag \\
M_{a}M_{\beta } &=&M_{\beta }M_{\alpha },\qquad \text{for\quad }\left\vert
\alpha -\beta \right\vert \geq 2,
\end{eqnarray}%
by setting%
\begin{equation}
\mathcal{B}_{\alpha }^{\left( 4\right) }=\frac{1}{\sqrt{2}}\left(
1+M_{\alpha }\right) .
\end{equation}%
It is straightforward to show that 
\begin{eqnarray}
\left( M_{\alpha }^{\left( 4\right) }\right) ^{2} &=&-1,  \notag \\
M_{\alpha }^{\left( 4\right) }M_{\alpha +1}^{\left( 4\right) } &=&-M_{\alpha
+1}^{\left( 4\right) }M_{\alpha }^{\left( 4\right) },  \notag \\
M_{\alpha }^{\left( 4\right) }M_{\alpha +2}^{\left( 4\right) } &=&M_{\alpha
+2}^{\left( 4\right) }M_{\alpha }^{\left( 4\right) },  \notag \\
M_{\alpha }^{\left( 4\right) }M_{\alpha +3}^{\left( 4\right) } &=&-M_{\alpha
+3}^{\left( 4\right) }M_{\alpha }^{\left( 4\right) },  \notag \\
M_{a}^{\left( 4\right) }M_{\beta }^{\left( 4\right) } &=&M_{\beta }^{\left(
4\right) }M_{\alpha }^{\left( 4\right) }\quad \text{for}\quad \left\vert
\alpha -\beta \right\vert \geq 4,
\end{eqnarray}%
when we set%
\begin{equation}
M_{\alpha }^{\left( 4\right) }\equiv i\gamma _{4}\gamma _{3}\gamma
_{2}\gamma _{1}.
\end{equation}%
It is a generalization of the extraspecial 2 group. Correspondingly, we
obtain a generalized braiding group relation%
\begin{equation}
\mathcal{B}_{\alpha }^{\left( 4\right) }\mathcal{B}_{\alpha +1}^{\left(
4\right) }\mathcal{B}_{\alpha }^{\left( 4\right) }=\mathcal{B}_{\alpha
+1}^{\left( 4\right) }\mathcal{B}_{\alpha }^{\left( 4\right) }\mathcal{B}%
_{\alpha +1}^{\left( 4\right) },\qquad \mathcal{B}_{\alpha }^{\left(
4\right) }\mathcal{B}_{\alpha +3}^{\left( 4\right) }\mathcal{B}_{\alpha
}^{\left( 4\right) }=\mathcal{B}_{\alpha +3}^{\left( 4\right) }\mathcal{B}%
_{\alpha }^{\left( 4\right) }\mathcal{B}_{\alpha +3}^{\left( 4\right) },
\end{equation}%
and%
\begin{equation}
\mathcal{B}_{\alpha }^{\left( 4\right) }\mathcal{B}_{\beta }^{\left(
4\right) }=\mathcal{B}_{\beta }^{\left( 4\right) }\mathcal{B}_{\alpha
}^{\left( 4\right) }
\end{equation}%
for $\left\vert \alpha -\beta \right\vert =2$ and $\left\vert \alpha -\beta
\right\vert \geq 4$. In the similar way, we find%
\begin{eqnarray}
\left( M_{\alpha }^{\left( 2N\right) }\right) ^{2} &=&-1,\qquad M_{\alpha
}^{\left( 4\right) }M_{\alpha +2n-1}^{\left( 4\right) }=-M_{\alpha
+2n-1}^{\left( 4\right) }M_{\alpha }^{\left( 4\right) },\qquad M_{\alpha
}^{\left( 4\right) }M_{\alpha +2n}^{\left( 4\right) }=M_{\alpha +2n}^{\left(
4\right) }M_{\alpha }^{\left( 4\right) },  \notag \\
M_{a}^{\left( 4\right) }M_{\beta }^{\left( 4\right) } &=&M_{\beta }^{\left(
4\right) }M_{\alpha }^{\left( 4\right) },\quad \text{for}\quad \left\vert
\alpha -\beta \right\vert \geq 2N
\end{eqnarray}%
for $1\leq n\leq N$. Hence, the $2N$-body Majorana operators satisfy a
genelized braiding group relation.%
\begin{equation}
\mathcal{B}_{\alpha }^{\left( 2N\right) }\mathcal{B}_{\alpha +2n-1}^{\left(
2N\right) }\mathcal{B}_{\alpha }^{\left( 2N\right) }=\mathcal{B}_{\alpha
+2n-1}^{\left( 2N\right) }\mathcal{B}_{\alpha }^{\left( 2N\right) }\mathcal{B%
}_{\alpha +2n-1}^{\left( 2N\right) },
\end{equation}%
and%
\begin{equation}
\mathcal{B}_{\alpha }^{\left( 2N\right) }\mathcal{B}_{\beta }^{\left(
2N\right) }=\mathcal{B}_{\beta }^{\left( 2N\right) }\mathcal{B}_{\alpha
}^{\left( 2N\right) }\quad \text{for\quad }\left\vert \alpha -\beta
\right\vert =2n\text{ and }\left\vert \alpha -\beta \right\vert \geq 2N,
\end{equation}%
for $1\leq n\leq N$.

\end{document}